\documentclass[aps,prb,twocolumn,superscriptaddress,showpacs,showkeys]{revtex4-1}

\usepackage{float}
\usepackage{latexsym}
\usepackage{mathtools}
\usepackage{amssymb}
\usepackage{amsmath}
\usepackage{epsfig}
\usepackage{graphicx} 
\usepackage{balance}
\usepackage{braket}
\usepackage{amsfonts}
\usepackage{empheq}
\usepackage{enumitem}
\usepackage{lipsum}

\usepackage[font=small, justification=raggedright, singlelinecheck=false, format=plain]{caption}
\usepackage[font=small, justification=justified, singlelinecheck=false, format=plain]{subcaption}
\usepackage{float}
\usepackage{dsfont}  
\usepackage{wasysym}
\usepackage{stackengine}

\usepackage{color}
\usepackage[dvipsnames]{xcolor}

\newlength{\upit}\upit=0.1truein

\newcommand{\ltappr}{{{\lower4pthbox{$<$} } \atop \widetilde{ \ \ \ }}}
\newlength{\bxwidth}\bxwidth=1.5 truein

\newcommand{\beq}{\begin{equation}}
\newcommand{\eeq}{\end{equation}}
\newcommand \bea {\begin{eqnarray} }
\newcommand \eea {\end{eqnarray}}

\newcommand{\ti}{$\times$} 
\newcommand{\vect}[1]{\boldsymbol{#1}}

\newcommand{\m}{$\mathds{1}_2$}

\newcommand\numberthis{\addtocounter{equation}{1}\tag{\theequation}}

\begin{document} 

\title{Symmetric spin liquids on the stuffed honeycomb lattice}

\author{Jyotisman Sahoo}
\affiliation{Department of Physics and Astronomy, Iowa State University, USA}
\author{Rebecca Flint}
\email{flint@iastate.edu}
\affiliation{Department of Physics and Astronomy, Iowa State University, USA}

\begin{abstract}
We use a projective symmetry group analysis to determine all symmetric spin liquids on the stuffed honeycomb lattice Heisenberg model. This lattice interpolates between honeycomb, triangular and dice lattices, always preserving hexagonal symmetry, and it already has one spin liquid candidate, TbInO$_3$, albeit with strong spin-orbit coupling not considered here.  In addition to the stuffed honeycomb lattice itself, we gain valuable insight into potential spin liquids on the honeycomb and triangular lattices, as well as how they might be connected.  For example, the sublattice pairing state proposed on the honeycomb lattice connects to the uniform spinon Fermi surface that may be relevant for the triangular lattice with ring exchange, while there are no spin liquids competitive on both the $J_1-J_2$ honeycomb and triangular lattice limits. In particular, we find three stuffed honeycomb descendants of the U(1) Dirac spin liquid widely believed to be found on the $J_1-J_2$ triangular lattice.  We also discuss how spin liquids near the honeycomb limit can potentially explain the physics of LiZn$_2$Mo$_3$O$_8$. 
\end{abstract}
\maketitle

\section{Introduction}

Quantum spin liquids provide an exciting and relatively simple example of topological order with low energy fractional spin 1/2 excitations and emergent gauge fluctuations\cite{savary16}.  Unfortunately, it is difficult to find quantum spin liquids either in materials or in models. Magnetic frustration plays a key role in stabilizing spin liquids over competing orders.  While the kagome lattice provides the best current candidates \cite{yan11,ran07,wang06,sachdev92,liao17,norman16}, there are also potential spin liquids on both the next-nearest-neighbor honeycomb \cite{gong13,clark11,albuquerque11,ganesh13,zhu13,ferrari17} and triangular \cite{kaneko14,shimada18,saadatmand15,itou08, wietek17,iqbal16,shijie19} lattices. Our paper addresses potential spin liquids on the frustrated \emph{stuffed honeycomb} lattice\cite{js18} that interpolates between the triangular and honeycomb lattices.  Our analysis sheds light on how the spin liquids might evolve between the two and provides guidance for future numerical calculations.

\begin{figure}[htbp]
\begin{center}
\includegraphics[scale = 0.15]{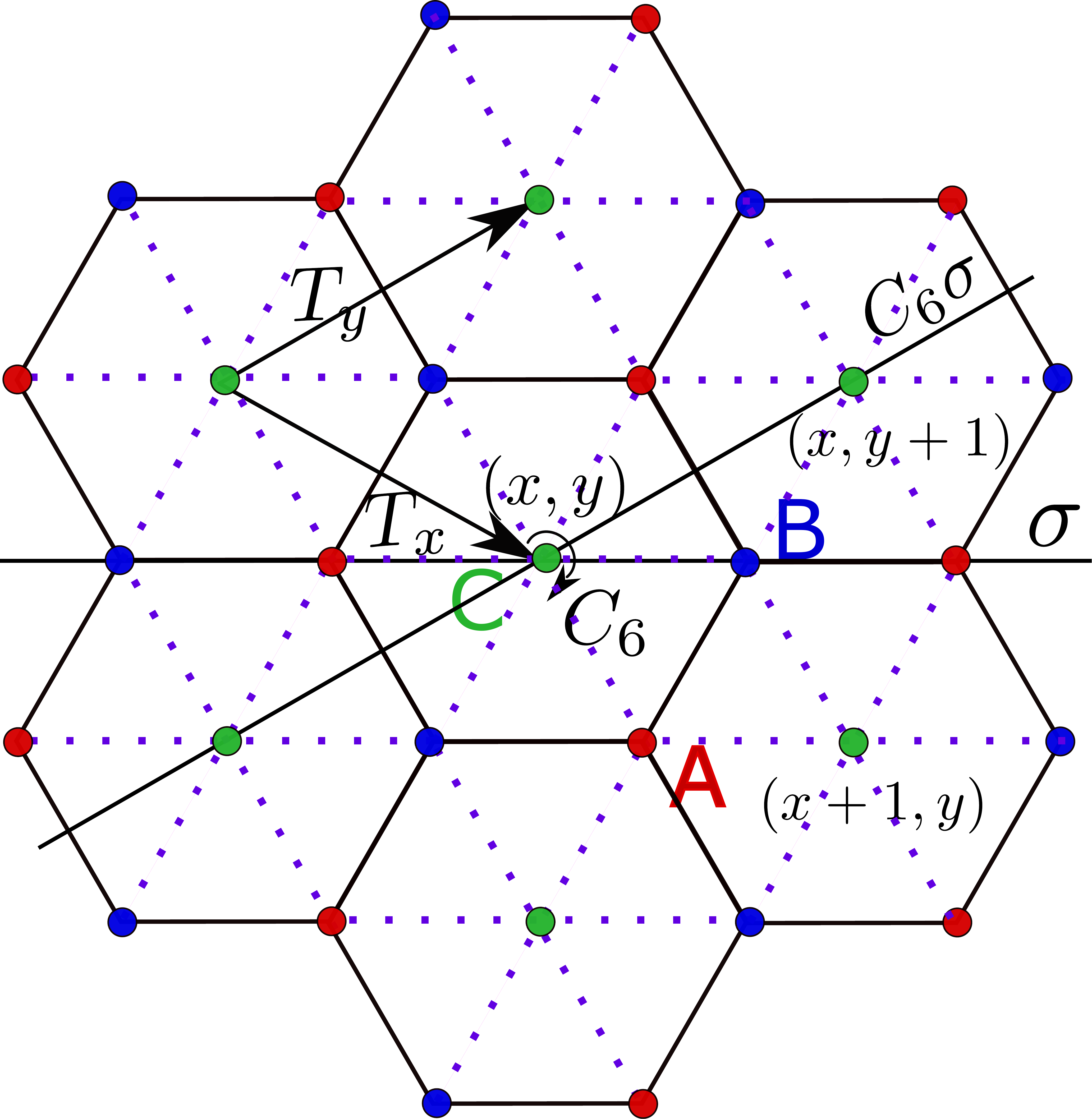}
\end{center}
\caption{\label{sgelem}{{\bf Model}: This figure shows the stuffed honeycomb lattice with three sublattices indicated as A/B (blue/green) and C (red).  A and B are related to each other by symmetry, but not to C.  Each site is represented as (x,y,s) with (x,y) giving the Bravais lattice site index and `s' denoting the sublattice. The solid and dotted bonds indicate the two different nearest-neighbor couplings ($J_1$ and $J'$). The space group generators are also shown: $T_{\hat{x}}$ and $T_{\hat{y}}$ are translations, while $\sigma$ and $C_6\sigma$ denote mirror planes, and $C_6$ is a six-fold rotation axis about the C sites}.}
\end{figure}

The stuffed honeycomb lattice is a non-Bravais lattice with space group {\it p6m}, containing three sublattices (ABC) in a hexagonal unit cell. It can be thought of as a honeycomb lattice (AB) with additional spins (C) at the center of each hexagon, effectively coupling honeycomb and triangular lattices.  The lattice and its space group generators are shown in Fig.\ref{sgelem}.  Note that the A and B sites are related by symmetry while C is symmetry related only in the triangular lattice limit. There are two types of nearest-neighbor bonds not related by symmetry, which we call $J_1$ for the bonds on the honeycomb sublattices and $J'$ for bonds between the C and honeycomb spins. This model interpolates from the honeycomb (decoupled from a C spin triangular lattice) at $J' = 0$ to the triangular for $J' = J_1$, both of which potentially host spin liquid regions\cite{gong13,clark11,albuquerque11,ganesh13,zhu13,ferrari17,kaneko14,shimada18,saadatmand15,itou08, wietek17,iqbal16,shijie19}, and the dice lattice for $J_1 = 0$, all the while maintaining the hexagonal symmetry, in contrast to the usual anisotropic triangular lattices\cite{coldea01,shimizu03,yamashita08,itou08,chubukov13,starykh15,mckenzie98}.  The classical phase diagram of this model was found in Ref. \onlinecite{js18}, while previous work has also examined a region of partial (C sublattice) disorder near the honeycomb limit with only nearest neighbor couplings\cite{nakano17,shimada18,gonzalez18,shimada182,siefert19}; further neighbor couplings will remove this region of partial disorder as they induce independent C spin order.

This lattice has been realized in rare earth indium oxides, RInO$_3$, R = (Gd,Tb,Dy) \cite{gordon18,peng18}. TbInO$_3$ is particularly interesting\cite{clark19,kim19,kimprb19}, as it does not order down to the lowest experimentally accessible temperatures; while Tb is a non-Kramers ion, both Tb sites seem to have magnetic moments and inelastic neutron scattering finds a low energy continuum reminiscent of spinons.

This model was originally introduced\cite{flint13} to explain the magnetic behavior of the cluster magnet LiZn$_2$Mo$_3$O$_8$ \cite{sheckelton12,mourigal14,sheckelton14,sheckelton15,chen16,carrasquilla17,chen18}. In this material, Mo$_3$O$_{13}$ molecular clusters carry spin $\frac{1}{2}$ and sit on a triangular lattice.  The high temperature Curie-Weiss susceptibility reflects all spins, but two-thirds of them vanish below 100K.  It has been proposed that a spontaneous lattice symmetry breaking leads to an emergent honeycomb lattice hosting a valence bond solid or spin liquid\cite{sheckelton12,flint13}, weakly coupled to the remaining one-third of the spins, which remain disordered down to low temperatures; hence LiZn$_2$Mo$_3$O$_8$ could potentially exist near the honeycomb limit of the stuffed honeycomb phase diagram.

In this paper, we determine all symmetric $\mathds{Z}_2$ spin liquids for Heisenberg spins on the stuffed honeycomb lattice using a projective symmetry group (PSG)\cite{wen02} analysis. This analysis also provides a convenient way to choose variational wavefunctions in a variational Monte Carlo (VMC) simulation\cite{ghorbani16}. This kind of analysis has already been done for the triangular\cite{bieri16,lu16} and honeycomb\cite{lu11}  limits, and here we show how those two limits may be connected, as well as finding spin liquids present only in the full stuffed honeycomb lattice.

The organization of the paper is as follows.  We review the basics of PSG analysis in Sec.\ref{bkg}, and apply this analysis to the stuffed honeycomb lattice in Sec.\ref{psgSH}. In Sec.\ref{const}, we show how the spin liquid {\it ansatze} are determined from PSGs. Our results are shown in Sec.\ref{result}, including a discussion of spin liquids previously found in the honeycomb and triangular limits, as well as the application to LiZn$_2$Mo$_3$O$_8$.  Sec. \ref{Tlim} discusses the three descendants of the U(1) Dirac spin liquid found on the stuffed honeycomb, while Sec. \ref{concl} summarizes our results.

\section{Background}\label{bkg}

Projective symmetry group (PSG) analysis is used to classify the possible spin liquid phases on a given lattice. The analysis can be done using either a fermionic or bosonic representation for the spins, which naturally captures spinons in the spin liquid phases. Here we restrict ourselves to the more general fermionic representation.  It naturally has an emergent SU(2) gauge freedom, which complicates and enriches any symmetry analysis, requiring the use of \emph{projective} symmetry groups that effectively project out the gauge symmetry.  The gauge symmetry means that apparently different states may actually be related by a gauge transformation. And a global symmetry operation, $\mathcal{R}$ (like rotation, translation, etc.) may not leave a symmetric state invariant, as it must be followed by an appropriate gauge transformation, $g_{\mathcal{R}}$ to reveal the full symmetry of the state. These combined operations, $(g_{\mathcal{R}},\mathcal{R})$ form the projective symmetry group\cite{wen02}.

In principle, any set of $\{(\mathcal{R},G_{\mathcal{R}})\}$ gives a different projective symmetry group and thus a different spin liquid.  However, in practice the allowed combinations of gauge transformations and symmetries are greatly constrained by the algebraic relations that the space group must satisfy.  Furthermore, any of the remaining allowed combinations are gauge-equivalent to one another; once this redundancy has been removed, there is usually a small set of distinct PSGs that lead to distinct spin liquids.  The properties of these spin liquids can then be analyzed in the fermionic mean-field theory allowed by the projective symmetries, and beyond.  

We consider the generic Heisenberg Hamiltonian:
\beq
H = \sum_{\left<ij\right> }J_{ij}\,\, {\bf S}_i\cdot{\bf S}_j.
\eeq
The spin operators can be expressed as
\beq\label{fermrep}
S_i^a = \frac{1}{2}{f}_{i\alpha}^{\dagger}\sigma_a^{\alpha\beta}{f}_{i\beta},
\eeq
where we use Einstein summation, `i' denotes the site index and $\sigma_a$ are the Pauli matrices. $f_{i\alpha}$ creates a neutral fermion of spin $\alpha = \uparrow or \downarrow$ at site `i'. This representation is invariant under the continuous SU(2) particle-hole transformation : $f_\alpha \mapsto \cos\theta f_\alpha + \sin\theta f_{\overline{\alpha}}^{\dagger}$, where $\overline\alpha = -\alpha$. While here we consider a Heisenberg Hamiltonian, the PSGs that emerge from this analysis are much more general and will represent any model with the same $SU(2)$, lattice and time-reversal symmetries, including higher order spin terms like ring exchange.

Although the spin commutation relations are satisfied by both sides of eq.\eqref{fermrep}, the dimensions of the Hilbert spaces do not match. The spin space per site is two-dimensional, while the fermionic Hilbert space per site is four-dimensional. To restrict our description to the physical spin-1/2 subspace, we must introduce the following constraint of one fermion per site,
\beq\label{cons}
\sum_{\alpha}f_{i\alpha}^{\dagger}f_{i\alpha} = 1,\,\,\,f_{i\alpha}f_{i\overline{\alpha}} = 0.
\eeq
The redundancy of our fermionic Hilbert space leads to an SU(2) gauge redundancy that the original spin space did not have. We introduce the Nambu spinor ${\psi}_i = (f_{i\uparrow},f_{i\downarrow}^{\dagger})$, where a local SU(2) transformation: $W_i\psi_i$ [$W_i$ $\in$ SU(2)] leaves the spin operator invariant. 

Inserting this fermionic representation into the Heisenberg Hamiltonian leads to a quartic fermionic Hamiltonian that may be decoupled with the mean-field amplitudes $\xi_{ij} = \left<f_{i\alpha}^{\dagger}f_{j\alpha}\right> $ (`hopping') and $\Delta_{ij} = \left<f_{i\uparrow}f_{j\downarrow}\right>$ (`pairing')\cite{wen02}. This mean-field theory becomes exact if the SU(2) spins are generalized to Sp(N) spins\cite{flint12}. 
In the mean-field picture, the exact constraints become
\beq
\left<f_{i\alpha}^{\dagger}f_{i\alpha}\right> = 1,\,\,\,\left<f_{i\alpha}f_{i\overline{\alpha}}\right> = 0.
\eeq 
 These conditions are enforced by Lagrange multipliers, $\lambda_j^a$, which adds the following term to the Hamiltonian: $$\sum_{j}\{\lambda_j^3(f_{j\alpha}^{\dagger}f_{j\alpha}-1) + [(\lambda_j^1+ i\lambda_j^2) f_{j\downarrow}f_{j\uparrow}+ H.c.]\}$$ 
The resulting quadratic mean-field Hamiltonian can be rewritten in terms of the Nambu spinor $\psi_i$,
\beq\label{mfh}
H_{MF} = \sum_{i,j}{\psi}_i^{\dagger}u_{ij}{\psi}_j + {\text H.c.} + \sum_{j}\lambda_j^a{\psi}_j^{\dagger}\sigma_a{\psi}_j
\eeq 
where the $2\times2$  link matrices $u_{ij}$ are:
\begin{equation}\label{ulabel}
 u_{ij} = \begin{pmatrix}
    \xi_{ij}       & \Delta_{ij} \\
    \Delta_{ij}^*       & -\xi_{ij}^*
\end{pmatrix}.
\end{equation}
These may be compactly written as $u_{ij} = u_{ij}^{\mu}\tau_\mu$ where $\{\tau_\mu\}$ are the Pauli matrices in Nambu space, (i\(\mathds{1}\)$_2$, $\tau_a$), $a = 1,2,3$ and $[u_{ij}]^{\dagger} = u_{ji}$. The parameters $u_{ij}^{\mu}$ can be written as $\{u_{ij}^{\mu}\}$ = $\{\xi_{ij}^{(2)},\Delta_{ij}^{(1)},\Delta_{ij}^{(2)},\xi_{ij}^{(1)}\}$, where $\xi_{ij} = \xi_{ij}^{(1)} + i\xi_{ij}^{(2)}$ are complex hopping and $\Delta_{ij} = \Delta_{ij}^{(1)} + i\Delta_{ij}^{(2)}$ are complex singlet pairing amplitudes.  Similarly, we can treat the Lagrange multipliers as the matrices $\lambda_j = \lambda_j^a\tau_a$.

The eigenvalues of \eqref{mfh} give the single spinon dispersion for a given mean-field {\it ansatz}. The physical symmetries may not be respected in the single spinon spectrum, as only the two-spinon continuum is physical. However, the dispersion can still tell us a lot about the nature of the spin liquid, particularly its low energy structure.  

The spinons only capture half the story; there are also emergent gauge fluctuations. Fixing $u_{ij}$ and $\lambda_j$ typically breaks the local SU(2) gauge freedom. However, no {\it ansatz} can completely break it and $H_{MF}$ may still be gauge invariant under a SU(2), U(1) or $\mathds{Z}_2$ transformation. These spin liquids are called SU(2), U(1) or $\mathds{Z}_2$ spin liquids. There always exists a subgroup, $\{W_i\}$ of SU(2) gauge transformations that leave the {\it ansatz} unchanged,
\beq\label{gt}
u_{ij} = W_i^{\dagger}u_{ij}W_j\,\,\forall\,\, u_{ij}.
\eeq
$\{W_i\} = \{\pm\mathds{1}_2\}$ is the minimal group that preserves the {\it ansatz}, as
any $u_{ij}$ is invariant under a global $\mathds{Z}_2$ transformation. Generically, this remaining gauge freedom is the invariant gauge group (IGG)\cite{wen02}, and helps determine the gauge/symmetry combinations. We will focus on $\mathds{Z}_2$ spin liquids, where the gauge fluctuations are gapped and the mean-field solution is stable. Here, gauge fluctuations are weak, but fluctuations are important for U(1) or SU(2) spin liquids, where they are not gapped. In such cases the mean-field theory may not be stable and the fermions and gauge fields are strongly coupled\cite{wen02,hermele04}.

The goal of our PSG analysis is to find the allowed combinations of gauge and symmetry operations, $\{(\mathcal{R},G_{\mathcal{R}})\}$ and their associated mean-field {\it ansatze} that respect all the symmetries of the lattice, once the gauge degree of freedom is considered.  The Hamiltonian has SU(2) spin rotation symmetry, time-reversal and space group symmetries.  The SU(2) spin symmetry is automatically satisfied by $H_{MF}$ if  the $u_{ij}^\mu$'s are real\cite{bieri16}. In the rest of the paper, we consider only the discrete symmetries.



There are a number of previous PSG analyses\cite{wen02,lu11,bieri16,lu16}, with different notation; here, we mostly follow that of Bieri, Lhullier and Messio \cite{bieri16}.  Within our mean-field picture, a given PSG will define a spin liquid phase described by a collection of {\it ansatze} $u_{ij}$ that remain invariant under a particular combination of the symmetry operations and SU(2) gauge transformations.  The space group (SG) and time-reversal symmetry elements ($\mathcal{R}$) will thus be accompanied by spatially dependent SU(2) gauge transformations ($g_\mathcal{R}$) to form the elements of the PSG, $Q_{\mathcal{R}} = (g_{\mathcal{R}},\mathcal{R})$ $\in$ SG$\ltimes \mathcal{G}$. These elements act on an {\it ansatz} $u = [u_{ij};\lambda_j]$ as
\beq\label{psgdef}
Q_\mathcal{R}(u) = [g_{\mathcal{R}}(i)u_{\mathcal{R}^{-1}(i,j)}g_{\mathcal{R}}^{\dagger}(j); g_{\mathcal{R}}(j)\lambda_{\mathcal{R}^{-1}(j)}g_{\mathcal{R}}^{\dagger}(j)].
\eeq
From this, we find the multiplication rule,
\beq\label{psg_mult}
Q_AQ_B = (g_A, A)(g_B, B) = (g_AAg_BA^{-1}, AB).
\eeq
An ansatz allowed by the PSG will satisfy,
\beq\label{invA}
Q_\mathcal{R}(u) = u.
\eeq
As any element, $W_i =\pm \mathds{1}_2$ in the IGG leaves all {\it ansatz} invariant, these are all possible gauge elements associated with the space group identity  ${\bf e}$.  Any  algebraic relationship in the SG only needs to be respected in the PSG up to the IGG, meaning up to a sign for $\mathds{Z}_2$ spin liquids,
\beq\label{Ig}
Q_xQ_y = \omega(x,y)Q_{xy},
\eeq
where $\omega(x,y) = \pm 1$. These signs are important to distinguish different PSGs.  The $g_{\mathcal{R}}$'s associated with the symmetry elements are not unique, as a gauge transformation $W$ can take us from one representation of the PSG to another via the transformation,
\beq\label{gt3}
g_{\mathcal{R}}(i) \mapsto W_i^{\dagger}g_{\mathcal{R}}(i)W_{\mathcal{R}^{-1}(i)}.
\eeq

To summarize, for any symmetry transformation, $\mathcal{R}$, the projective symmetry transformation is $Q_{\mathcal{R}} = (g_{\mathcal{R}},\mathcal{R})$. These $Q_{\mathcal{R}}$'s leave the allowed mean-field {\it ansatze} invariant, and can be constrained by requiring the $Q_{\mathcal{R}}$'s to satisfy the algebraic relations of the space group, up to the IGG. There are actually equivalence classes of $\{Q_\mathcal{R}\}$ that can be related by local gauge transformations; we will proceed to find an example of each equivalence class on the stuffed honeycomb lattice.

\section{PSGs of the stuffed honeycomb}\label{psgSH}

In this section, we develop the distinct PSGs for the stuffed honeycomb lattice by determining the $Q_{\mathcal{R}}$'s satisfying the algebraic relations followed by the space group.

\subsection{Space group elements}

The space group (SG) of the stuffed honeycomb lattice is generated by: SG = $\{T_{\hat{x}}, T_{\hat{y}}, C_6, \sigma\}$. $T_{\hat{x}}$ and $T_{\hat{y}}$ are translation operators, shown in Fig. \ref{sgelem}, while $C_6$ and $\sigma$ are six-fold rotation and reflection operators, respectively. These generators obey the algebraic relations:
\begin{subequations}\label{alg1}
\begin{align}
T_{\hat{x}} T_{\hat{y}} = T_{\hat{y}} T_{\hat{x}} \label{rel1}\\ 
\sigma T_{\hat{x}} = T_{\hat{y}} \sigma \label{rel2}\\
T_{\hat{y}} C_6 = C_6 T_{\hat{x}} = T_{\hat{x}} C_6 T_{\hat{y}} \label{rel3}
\end{align}
\end{subequations}
and
\begin{subequations}\label{alg2}
\begin{align}
\sigma^2 = e \label{rel5}\\ 
C_6^6 = e \label{rel6}\\
(C_6 \sigma)^2 = e \label{rel8}
\end{align}
\end{subequations}
These are equivalent to those of the honeycomb lattice\cite{lu11}, as the space groups are identical. 

Now we determine the gauge representations, $g_{\mathcal{R}}$ associated with each $\mathcal{R}\in$ SG. In general, these are spatially dependent (i), depending both on the unit cell [(x,y)] and sublattice (s=A,B,C). We start with the translation operators, where the first algebraic relation, \eqref{rel1} requires
\beq\label{translation}
g_{\hat{x}}(i)T_{\hat{x}}g_{\hat{y}}(i)T_{\hat{x}}^{-1} = \epsilon_2 g_{\hat{y}}(i)T_{\hat{y}}g_{\hat{x}}(i)T_{\hat{y}}^{-1}.
\eeq
`i' indicates $(x,y,s)$ and $\epsilon_2$ is a shorthand for the sign $\omega(T_{\hat{x}},T_{\hat{y}})\omega(T_{\hat{y}},T_{\hat{x}})^{-1}$. We can use eq. \eqref{gt3} to gauge fix $g_{\hat{x}}$ to be $\mathds{1}_2$, which simplifies \eqref{translation} to:
\beq
\begin{split}
&g_{\hat{y}}(x-1,y,s) = \epsilon_2 g_{\hat{y}}(x,y,s)\\
&g_{\hat{y}}(x,y,s) = \epsilon_2^x g_{\hat{y}}(y,s).
\end{split}
\eeq
We can simultaneously remove the y-dependence of $g_{\hat y}$ by using a y-dependent gauge transformation to gauge fix  $g_{\hat{y}}$; this gauge transformation leaves $g_{\hat x}$ invariant, as it must.  Now we have the representations for both translation operators,
\beq\label{T}
\begin{split}
&g_{\hat{x}}(x,y,s) =  \mathds{1}_2\\
&g_{\hat{y}}(x,y,s) = (\epsilon_{2})^x \mathds{1}_2. 
\end{split}
\eeq
The rotation and reflection operators are treated similarly, as shown in the appendix. While several other signs, like $\epsilon_2$ appear, these can all be removed by further gauge fixing that continues to leave all previously determined $g_\mathcal{R}$'s unchanged up to a sign. The final set of gauge representations is,
\begin{equation}\label{PSG1}
\begin{aligned}
&g_\sigma(x,y,A) = (\epsilon_2)^{x(y+1)} g_{\sigma,A}\\
&g_\sigma(x,y,B) = (\epsilon_2)^{xy} g_{\sigma,B}\\
&g_\sigma(x,y,C) = (\epsilon_{2})^{xy} g_{\sigma,C}\\
&g_{C_6}(x,y,A) =  (\epsilon_{2})^{(x+1)y + x(x-1)/2}g_{R,A}\\
&g_{C_6}(x,y,B) =  (\epsilon_{2})^{(x+1)y + x(x-1)/2}g_{R,B}\\
&g_{C_6}(x,y,C) =  (\epsilon_{2})^{(x+1)y + x(x-1)/2}g_{R,C}.
\end{aligned}
\end{equation}
Here, $g_{\sigma,s}$ and $g_{R,s}$ are all uniform SU(2) matrices. These are generically different for different sublattices, however we can find a gauge where $g_{R,B} = g_{R,A}$, but not so for $g_{\sigma,s}$. Consider the general sublattice gauge transformation, $g = (g_A, g_B, g_C)$. The $g_{R,s}$'s transform as,
\begin{equation}\label{gR}
\begin{aligned}
g_{R,A} \rightarrow g_Ag_{R,A}g_B^{\dagger}\\
g_{R,B} \rightarrow g_Bg_{R,B}g_A^{\dagger}\\
g_{R,C} \rightarrow g_Cg_{R,C}g_C^{\dagger}.
\end{aligned}
\end{equation}
We can then define a new matrix, $g_{R}$ by $g_R^2 = g_{R,B}g_{R,A}$. If we take the gauge $g = (g_{R,B},g_R,g_{R,C})$, then
\begin{equation}\label{gR2}
(g_{R,A},g_{R,B},g_{R,C}) \mapsto (g_R,g_R,g_{R,C}).
\end{equation}
Hence, in our gauge, $g_{R,A} = g_{R,B} = g_{R}$. At this point, we have only used eq. \eqref{alg1}. 
The $g_{\sigma,s}$'s and $g_{R,s}$'s are further constrained by \eqref{alg2}:
\begin{equation}\label{PSG2}
\begin{aligned}
(g_{\sigma,s})^2 = \epsilon_{\sigma}\mathds{1}_2\\
g_{R,C}^6 = g_{R}^6 = \epsilon_{R}\mathds{1}_2\\
(g_{R,C}g_{\sigma,C})^2 = \epsilon_{R\sigma}\mathds{1}_2\\
g_Rg_{\sigma,B}g_Rg_{\sigma,A} = g_Rg_{\sigma,A}g_Rg_{\sigma,B} =\epsilon_2\epsilon_{R\sigma}\mathds{1}_2 
\end{aligned}
\end{equation}
where $\epsilon_\sigma$, $\epsilon_R$, $\epsilon_{R\sigma}$ = $\pm 1$ are all combinations of different $\omega(\mathcal{R},\mathcal{R'})$ in the IGG. These are all the constraints on the PSGs given by the space group elements, but they will be further constrained by time-reversal symmetry.

\begin{table}[t]
\begin{center}
\resizebox{\columnwidth}{!}{%
\begin{tabular}{|l|l|l|l|l|l|l|l|l|l|l|l|l|l|l|l|l|l|l|l|l|l|l|}

\hline
          No. & $\epsilon_2$  &$g_{\sigma C}$ & $g_{\sigma B}$ &$g_{\sigma A}$& $g_{RC}$ & $g_R$ & $\lambda_B$ & $\lambda_C$ &$u_1$ & $u'$ & $u_C$&$u_2$&gap&flux\\
\hline
\hline
1 &-    &i$\tau_2$  &i$\tau_2$&i$\tau_2$  &i$\tau_2$  &i$\tau_3$  &2  &2 &2&2&\ti&2&G&($\pi$,0)\\
\hline

2 &-    &i$\tau_2$  &i$\tau_2$&-i$\tau_2$  & i$\tau_3$ & i$\tau_3$ & 2 &\ti &1&2&2&2&D&($\pi$,0)\\
\hline
3 &-   &i$\tau_2$  &i$\tau_2$&-i$\tau_2$  &i$\tau_3$  &b  &\ti  & \ti&1&2&2&2&D&($\pi$,0)\\
\hline
4  &+  &\m  &\m &\m &\m  &\m  &1,3  &1,3 &1,3 &1,3 &1,3 &1,3&G$^*$&(0,0)\\
\hline
4a  &+  &i$\tau_3$  &i$\tau_3$ &i$\tau_3$ &\m  &\m  &3  &3 &3 &3 &3 &3&G$^*$&(0,0)\\
\hline
5  &+    &i$\tau_1$  &i$\tau_1$&i$\tau_1$  &a  &a  &\ti  &\ti &1&1&1&1&Q,D&NC\\
\hline
6 &+   &\m  &\m &\m  &i$\tau_3$  &i$\tau_3$  &1,3  &3 &3&1,3&3&1,3&G$^*$&(0,0)\\
\hline
6a &+   &i$\tau_3$  &i$\tau_3$ &i$\tau_3$  &i$\tau_3$  &i$\tau_3$  &3  &3 &3&3&3&3&G$^*$&(0,0)\\
\hline
7 & +  & i$\tau_1$ &i$\tau_1$&i$\tau_1$  &\m  &a  &\ti  &1 &1&1&1&1&D&(0,0)\\      
\hline
8  &+   &i$\tau_1$  &i$\tau_1$&i$\tau_1$  &a  &\m  &1  &\ti &1&1&1&1&Q&NC\\
\hline
9  &-     &i$\tau_2$  &i$\tau_2$&i$\tau_2$  &b  &i$\tau_2$  &2  &\ti &0&2&2&2&D&NC\\
\hline
10  &-   &i$\tau_2$  &i$\tau_2$&-i$\tau_2$  &b  &b  &\ti  &\ti &1&2&2&2&D&NC\\ 
\hline
\hline
\end{tabular}%
}
\end{center}
\caption{\label{QSL1}The table shows all PSGs with both nearest neighbor {\it ansatz} allowed. In the interest of space, we list only $\epsilon_2$, $g_{\sigma}$, and $g_{R}$, which are enough to identify a PSG and propagate the {\it ansatz}. While the $g_{\mathcal{R}}$'s are usually Pauli matrices, a$=\exp(i\tau_3\pi/3)$ and b$=\exp(i\tau_3\pi/6)$ are also allowed by the hexagonal symmetry for $g_R$,$g_{R,C}$. The ansatz allowed by the PSG are shown in the columns labeled with $\lambda^{\mu}$, $u^{\mu}$, which contain the allowed $\mu$'s with the values 0,1,2 or 3; $\times$ indicates that the ansatz is completely forbidden. The `gap' column describes the most generic single spinon dispersion, with the acronyms: `G'-gapped, `D'-Dirac, `FS'-spinon Fermi surface and `Q'-Quadratic band touching. There are two types of spin liquids that can be gapped out.  Some, labeled G, are gapped except for a measure zero region of the $u$ parameter space, while others, labeled G* have substantial fractions of parameter space where they are gapless (e.g. - PSG 4 may be gapped, or may realize a spinon Fermi surface). The `flux' column gives the SU(2) flux through the fundamental triangular plaquettes, calculated for each such plaquette around a given C site. While ($\pi$,0) denotes an alternating $\pi$ and 0 fluxes as shown in Fig.\ref{fluxfig}, (0,0) means a 0 flux throughout. `NC' stands for non-collinear flux, meaning that the directors vary between plaquettes.}
\end{table}

\begin{table}[t]
\begin{center}
\begin{tabular}{|l|l|l|l|l|l|l|l|l|l|l|l|l|l|l|l|l|l|l|l|l|l|}

\hline
          No. & $\epsilon_2$   &$g_{\sigma,C}$ & $g_{\sigma,B}$ &$g_{\sigma,A}$& $g_{R,C}$ & $g_R$ & $\lambda_B$ & $\lambda_C$ &$u_1$ & $u_C$&$u_2$&gap\\
\hline
\hline
1  &-    &i$\tau_3$  &i$\tau_2$&i$\tau_2$  &i$\tau_3$  &i$\tau_3$  &2  &3 &2&\ti&2&G\\
\hline
2  &+   &i$\tau_3$  &i$\tau_2$&-i$\tau_2$  &i$\tau_3$  &i$\tau_3$  &2  &3 &3&3&2&FS\\
\hline
3  &-   &i$\tau_3$  &i$\tau_2$&-i$\tau_2$  &i$\tau_1$  &i$\tau_1$  &2  &\ti &3&3&2&FS\\
\hline
4  &-   &i$\tau_1$  &i$\tau_2$&-i$\tau_2$  &b  &b  &\ti  &\ti &1&1&2&D\\
\hline
5  &-   &i$\tau_3$  &i$\tau_2$&-i$\tau_2$  &i$\tau_2$  &b  &\ti  &\ti &1&3&2&D\\
\hline
6  &-   &i$\tau_1$  &i$\tau_2$&-i$\tau_2$  &b  &i$\tau_3$  &2  &\ti &1&1&2&D\\
\hline
\hline
\end{tabular}
\end{center}
\caption{\label{QSL2}This table lists all PSGs without any allowed $u'$ {\it ansatz}; these are not expected to be good candidates near the triangular limit, but may be relevant near the honeycomb limit.  All notation is identical to table \ref{QSL1}.  There are no non-trivial nearest-neighbor triangle fluxes.}
\end{table}

\begin{table}[t]
\begin{center}
\begin{tabular}{|l|l|l|l|l|l|l|l|l|l|l|l|l|l|l|l|l|l|l|l|l|l|}

\hline
          No. & $\epsilon_2$   &$g_{\sigma,C}$ & $g_{\sigma,B}$ &$g_{\sigma,A}$& $g_{R,C}$ & $g_R$ & $\lambda_B$ & $\lambda_C$ & $u'$ & $u_C$&$u_2$&gap\\
\hline
\hline
1 &  -  & i$\tau_1$ & i$\tau_1$& i$\tau_1$ &i$\tau_3$  &i$\tau_1$  &1  &\ti &1&1&1&D\\
\hline
2  &+    &i$\tau_2$  &i$\tau_2$&i$\tau_2$  & i$\tau_3$ & i$\tau_3$ & 2 &\ti &2&\ti&2&FB\\
\hline
3  &+   &i$\tau_2$  &i$\tau_2$&i$\tau_2$  &b  &b  &\ti  &\ti &2&\ti&2&FB\\      
\hline
4  &-     &i$\tau_2$  &i$\tau_2$&i$\tau_2$  &i$\tau_2$  &b  &\ti  &2 &2&\ti&2&Q,D\\      
\hline
5 &+    &i$\tau_2$  &i$\tau_2$&i$\tau_2$  &b  &i$\tau_3$  &2  &\ti &2&\ti&2&FB\\
\hline
6  &+   &i$\tau_2$  &i$\tau_2$&i$\tau_2$  &i$\tau_3$  &b  &\ti  & \ti&2&\ti&2&FB\\
\hline
7  &-  &\m  &\m&-\m  &\m  &\m  &1,3  &1,3  &1,3 &\ti &1,3&G\\
\hline
7a  &- &i$\tau_3$  &i$\tau_3$&-i$\tau_3$  &\m  &\m  &3  &3 &3&\ti&3&G\\
\hline
8  &-   &i$\tau_1$  &i$\tau_1$&-i$\tau_1$  &a  &a  &\ti  &\ti &1&\ti&1&D\\
\hline
9  & + & i$\tau_1$ & i$\tau_1$& -i$\tau_1$ &i$\tau_3$  &i$\tau_1$  &1  &\ti &1&\ti&1&FB\\
\hline
10 &-    &\m  &\m &-\m  &i$\tau_3$  &i$\tau_3$  &1,3  &3 &1,3&1&1,3&G\\
\hline
11  &+    &i$\tau_2$  &i$\tau_2$&-i$\tau_2$  &i$\tau_2$  &b  &\ti  &2 &2&2&2&Q\\  
\hline
12 & -  & i$\tau_1$ &i$\tau_1$&-i$\tau_1$  &\m  &a  &\ti  &1&1&\ti&1&D\\   
\hline
13  &-   &i$\tau_1$  &i$\tau_1$&-i$\tau_1$  &a  &\m  &1  &\ti&1&\ti&1&Q,D\\
\hline 
14  &+    &i$\tau_2$  &i$\tau_2$&-i$\tau_2$  &b  &i$\tau_2$  &2  &\ti &2&\ti&2&FB\\
\hline
15 & - &i$\tau_1$ & i$\tau_1$& -i$\tau_1$ &i$\tau_2$  &b  &\ti  &\ti &1&1&1&Q,D\\
\hline
\hline
\end{tabular}
\end{center}
\caption{\label{QSL3}This table lists all PSGs without any allowed $u_1$ {\it ansatz}; these are not expected to be relevant near the triangular and honeycomb limits, but may be relevant near the dice limit.  Several of these have flat bands, indicated by `FB'.  These are unlikely to be stable mean-field phases.  All notation is again identical to table \ref{QSL1}.  There are again no non-trivial nearest-neighbor triangle fluxes.}
\end{table}

\subsection{Time Reversal}\label{TRcons}

Time-reversal acts on the spinor ${f}_i$ as $\mathcal{T}^{\dagger}{ f}_i\mathcal{T} = i\sigma_2{ f}_i$ or equivalently $\mathcal{T}^{\dagger}{\psi}_i\mathcal{T} = [(-i\tau_2{\psi}_i)^T]^{\dagger}$. It is convenient to use the gauge transformation: $\psi_i \mapsto i\tau_2\psi_i$ to reduce the time-reversal operation to
\beq\label{Tfunc}
\mathcal{T}: \psi_i\mapsto \psi_i^*.
\eeq  
This time-reversal operation inverts any generic {\it ansatz}, $u_{ij} \xrightarrow[]{\mathcal{T}} -u_{ij}$.  The $g_{\mathcal{R}}$'s are constrained by the commutation of $\mathcal{T}$ with all SG elements,
\begin{equation}\label{Tcom}
\begin{aligned}
&\mathcal{T}S = S\mathcal{T}\,\,\,\,\,\,\,\,\,\text{where $S = T_{\hat{x}}, T_{\hat{y}}, \sigma, C_6$}\\
&\mathcal{T}^2 = -1,
\end{aligned}
\end{equation}
where we derive these constraints in Appendix \ref{appTR}.
The first set of relations force $g_{\mathcal{T}}$ to be independent of unit cell, but allow for sublattice dependence ($g_{\mathcal{T},s}$) with the following conditions:
\begin{equation}\label{TR_Rs}
\begin{aligned}
&g_{R}g_{\mathcal{T},A} = \epsilon_{TR} g_{\mathcal{T},B}g_{R}\\
&g_{R}g_{\mathcal{T},B} = \epsilon_{TR} g_{\mathcal{T},A}g_{R}\\
&g_{R,C}g_{\mathcal{T},C} = \epsilon_{TR} g_{\mathcal{T},C}g_{R,C}\\
&g_{\sigma,s}g_{\mathcal{T},s} = \epsilon_{T\sigma} g_{\mathcal{T},s}g_{\sigma,s}\,\,\,\,\,\,\,\,\,\text{(s=A,B,C)},
\end{aligned}
\end{equation}
where $\epsilon_{TR}$, $\epsilon_{T\sigma}$ = $\pm 1$ are signs coming from the commutation relations from rotation and reflection operators, respectively. Finally, $\mathcal{T}^2 = -1$ requires that
\begin{equation}
g_{\mathcal{T},s}^2 =\epsilon_{\mathcal{T}}\mathds{1}_2.
\end{equation}
In general, the symmetry allowed {\it ansatz} and the time-reversal representation, $g_{\mathcal{T}}(i)$, must satisfy :
\beq\label{TRi}
-u_{ij} = g_{\mathcal{T}}(i) u_{ij} g_{\mathcal{T}}^{\dagger}(j).
\eeq
For uniform $g_{\mathcal{T}}$, this equation forbids imaginary hopping terms, (i$u^0 \mathds{1}_2$). Additionally, eq.\eqref{TRi} forces the {\it ansatz} components to be coplanar in this gauge. For example, if $g_{\mathcal{T}} = \tau_2$, then the {\it ansatze} are restricted to the $(\tau_1,\tau_3)$ plane.  Generically, $g_{\mathcal{T}}$ can be non-uniform which in principle allow a non-coplanar {\it ansatz}. In fact, we can convert a seemingly coplanar set into a non-coplanar one and vice-versa by a non-uniform gauge transformation.  In general, $\epsilon_{\mathcal{T}} = 1$ requires $g_{\mathcal{T},s} = \pm\mathds{1}_2$, which significantly restricts the ansatz such that any {\it ansatz} connecting symmetry related sites must vanish. In the tables that follow, we will consider only the case $\epsilon_{\mathcal{T}} = -1$.

\begin{figure*}[t]
\captionsetup[subfigure]{justification=centering}
\centering
\begin{subfigure}[t]{.28\textwidth}
 \centering
\includegraphics[width=\linewidth]{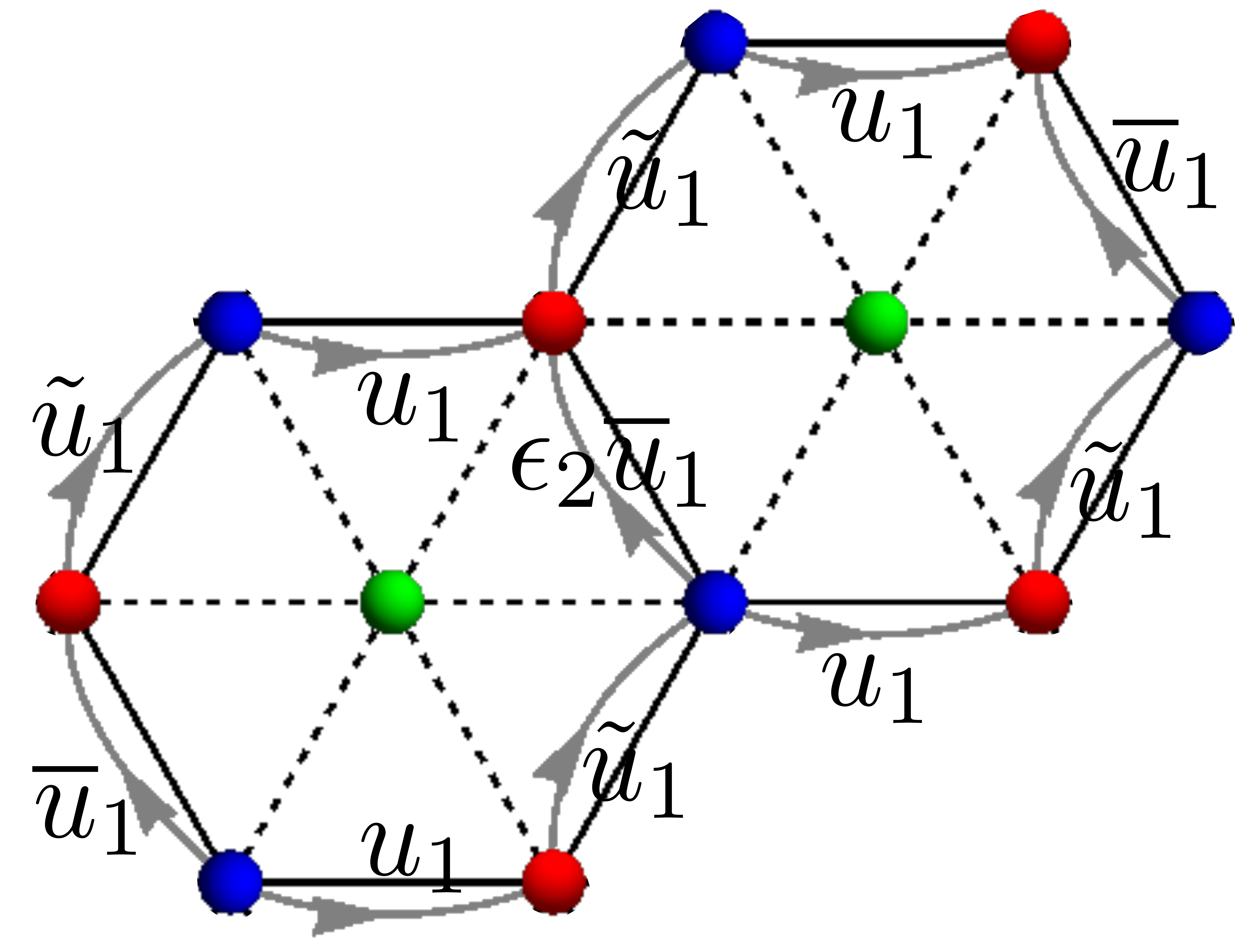}
  \label{p1}
  \caption{}
\end{subfigure}\hspace{10mm}
\begin{subfigure}[t]{.28\textwidth}
  \centering
\includegraphics[width=\linewidth]{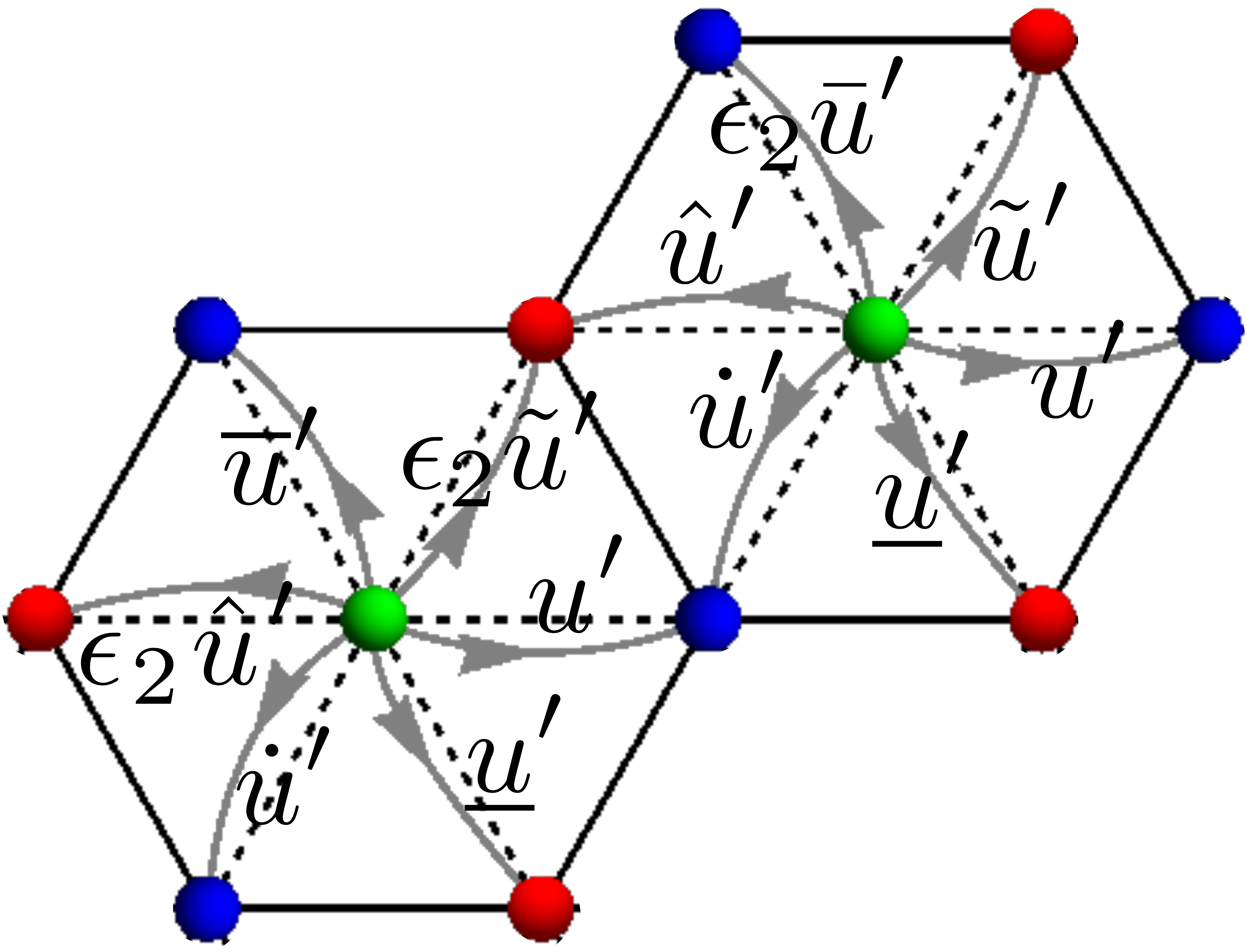}
  \label{p2}
  \caption{}
\end{subfigure}\\
\begin{subfigure}[t]{.28\textwidth}
  \centering
\includegraphics[width=\linewidth]{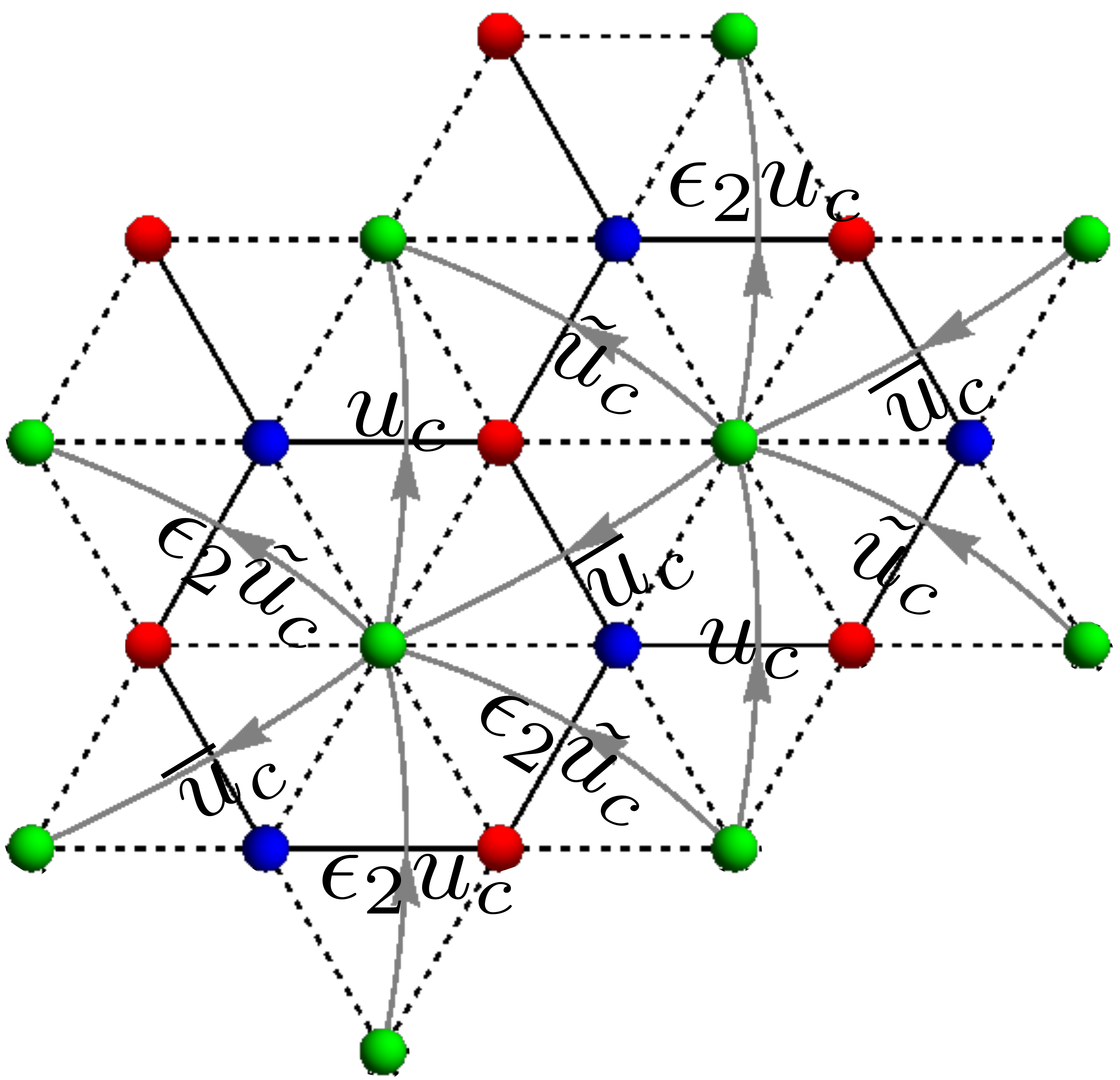}
  \label{int2}
  \caption{}
\end{subfigure}\hspace{10mm}
\begin{subfigure}[t]{.28\textwidth}
  \centering
\includegraphics[width=\linewidth]{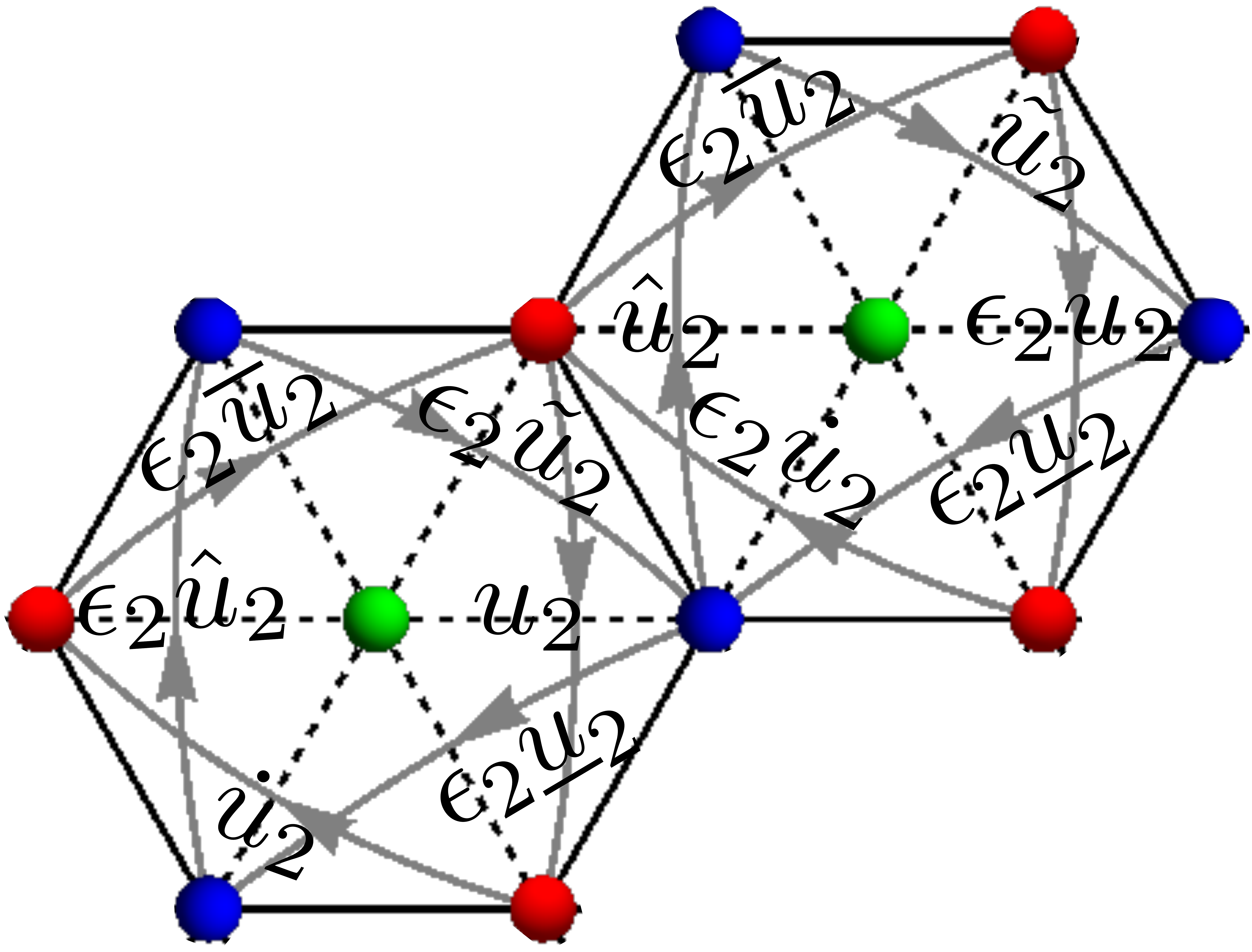}
  \label{int3}
  \caption{}
\end{subfigure}
\caption{\label{ansatzeprop} This figure shows how the {\it ansatze} may be propagated through the unit cell using the definitions in eqns.\eqref{propu1}-\eqref{propup}.  All {\it ansatze} can be captured within a six-site unit cell (doubled due to $\epsilon_2$). While (a) and (b) show the nearest neighbor {\it ansatze}, $u_1$ and $u'$, (c) and (d) show the same for the second neighbors, $u_C$ and $u_2$, respectively.}
\end{figure*}

\section{Spin liquid ansatze}\label{const}

A PSG is defined by the set of $\{\epsilon_2,\epsilon_R,\epsilon_{\sigma}, \epsilon_{R\sigma}\, \epsilon_{TR},\epsilon_{T\sigma}\}$ and $\{g_{\sigma,s},g_{R,C},g_R,g_{\mathcal{T},s}\}$. To understand the nature of the spin liquid, we need to find which {\it ansatze} are allowed by the PSG.  In order for an {\it ansatz} to have the full symmetry of the lattice, it needs to satisfy eq.\eqref{invA} for all symmetry operators. Explicitly,
\beq
\begin{split}
&g_{\mathcal{R}i}u_{\mathcal{R}^{-1}(ij)}g_{\mathcal{R}j}^{\dagger} = u_{ij}\\
&g_{\mathcal{R}j}\lambda_{\mathcal{R}^{-1}(j)}g_{\mathcal{R}j}^{\dagger} = \lambda_{j}.
\end{split}
\eeq
This requirement strongly constrains the {\it ansatze}. There are two sub-classes of constraints. The first is the case where $\mathcal{R}^{-1}(ij)$  = $(ij)$ or $(ji)$, which constrains the allowed $u_{ij}$. Similarly, if $\mathcal{R}^{-1}(j)$  = $j$, we constrain $\lambda_j$.  The second class relates different links/sites and gives the real space {\it ansatz}, which may appear to break translation symmetry and double the unit cell.

First, we enumerate the constraints.  $\lambda_C$ must respect both generators of the point group.
\begin{equation}\label{lc}
\begin{aligned}
\lambda_C = g_{\sigma,C}\lambda_C [g_{\sigma,C}]^{\dagger}\,\,\,\,\,\,\,\,\,\,\text{($\sigma$)}\\
\lambda_C = g_{R,C}\lambda_C [g_{R,C}]^{\dagger}\,\,\,\,\,\,\,\,\,\,\text{($R$)},
\end{aligned}
\end{equation}
while $\lambda_B$ is invariant under $\sigma$ and $T_{\hat{x}}R^2$, 
\begin{equation}\label{lb}
\begin{aligned}
\lambda_B = g_{\sigma,B}\lambda_B &[g_{\sigma,B}]^{\dagger}\,\,\,\,\,\,\,\,\,\,\text{($\sigma$)}\\
\lambda_B = g_{R}^2\lambda_B &[g_{R}^2]^{\dagger}\,\,\,\,\,\,\,\,\,\,\text{($T_{\hat{x}}R^2$)}.
\end{aligned}
\end{equation}
$\lambda_A$ is determined by $\lambda_A = g_R^{\dagger}\lambda_Bg_R$. 

The nearest-neighbor terms between AB sites ($u_1$) are constrained by $T_{\hat{y}}R^3T_{\hat{x}}$, which exchanges the sites, and $T_{\hat{x}}^{-1}T_{\hat{y}}\sigma$ which leaves the link invariant, 
\begin{subequations}\label{U1}
\begin{align}
u_1^{\dagger} = \epsilon_2g_R^3 u_1 (g_R^3)^{\dagger}\,\,\,\,\,\,\,\,\,\,\text{($T_{\hat{y}}R^3T_{\hat{x}}$)}\\
u_1 = (g_{\sigma,B}) u_1 (g_{\sigma,A})^{\dagger}\,\,\,\,\,\,\,\,\,\,\text{($T_{\hat{x}}^{-1}T_{\hat{y}}\sigma$)}
\end{align}
\end{subequations}

The nearest-neighbor terms connecting AC/BC ($u'$) are only constrained by $\sigma$, which leaves the links unchanged,
\begin{equation}\label{U2}
u' = g_{\sigma,C}u'g_{\sigma,B}^{\dagger}.\,\,\,\,\,\,\,\,\,\,\text{($\sigma$)}
\end{equation}

There are two different next-nearest neighbor terms that connect AA/BB sites ($u_2$) or CC sites ($u_C$). $u_2$ is only invariant under $\sigma$,
\begin{equation}\label{U3}
u_2^{\dagger} = (g_{\sigma,s})u_2 (g_{\sigma,s})^{\dagger}\,\,\,\,\,\,\,\,\,\,\text{($\sigma$)},
\end{equation} 
for $s = A,B$, while $u_C$ is constrained by
\begin{subequations}\label{U4}
\begin{align}
u_{C}^{\dagger} = \epsilon_{2}g_{R,C}^3 u_{C} (g_{R,C}^3)^{\dag}\,\,\,\,\,\,\,\,\,\,\text{($T_{\hat{y}}R^3T_{\hat{x}}$)}\\
u_{C}^{\dagger} = (g_{\sigma,C})u_C (g_{\sigma,C})^{\dagger}\,\,\,\,\,\,\,\,\,\,\text{($T_{\hat{x}}^{-1}T_{\hat{y}}\sigma$)}
\end{align}
\end{subequations}

These are all the constraints up to next-nearest-neighbor terms.  The second class of requirements relate the {\it ansatz} on different links/sites. This action propagates the fundamental $u_{ij}$'s determined above throughout the unit cell, as shown in Fig.\ref{ansatzeprop}.  As $\epsilon_2 = \pm 1$ allows translation to be broken along $\hat T_{\hat{y}}$, all {\it ansatz} can be captured within a six site unit cell. Finding how the {\it ansatz} propagates requires only the rotation and translation operators, and we label the transformed links,
\begin{equation}\label{propu1}
\begin{aligned}
&\tilde{u}_{1,2} = g_R u_{1,2} g_R^{\dagger}\\
&\bar{u}_{1,2} = g_R^2 u_{1,2} [g_R^2]^{\dagger}\\
&\hat{u}_2 = g_R^3 u_2 [g_R^3]^{\dagger}\\
&\dot{u}_2 = g_R^4 u_2 [g_R^4]^{\dagger}\\
&\underline{u}_2 = g_R^5 u_2 [g_R^5]^{\dagger}\\
&\lambda_A = g_R\lambda_B g_R^{\dagger}.
\end{aligned}
\end{equation}
For $u_C$, they have the following form,
\begin{equation}\label{propuc}
\begin{aligned}
&\tilde{u}_C = g_{R,C} u_C g_{R,C}^{\dagger}\\
&\bar{u}_C = g_{R,C}^2 u_C [g_{R,C}^2]^{\dagger}.
\end{aligned}
\end{equation}
Finally, for $u'$, we have
\begin{equation}\label{propup}
\begin{aligned}
&\tilde{u}' = g_{R,C} u' g_R^{\dagger}\\
&\bar{u}' = g_{R,C}^2 u' [g_R^2]^{\dagger}\\
&\hat{u}' = g_{R,C}^3 u' [g_R^3]^{\dagger}\\
&\dot{u}' = g_{R,C}^4 u' [g_R^4]^{\dagger}\\
&\underline{u}' = g_{R,C}^5 u' [g_R^5]^{\dagger}.
\end{aligned}
\end{equation}
Now we have the tools to transform any PSG into a symmetry allowed {\it ansatz} and thus obtain the spinon spectrum, which determines if the spinon spectrum is gapped or not, and the gauge fluxes through various plaquettes, which determine the allowed gauge fluctuations.


\section{Results}\label{result}

All unique symmetric spin liquids are tabulated in Tables \ref{QSL1}-\ref{QSL3}. The number has been greatly reduced both by the constraints of section \ref{psgSH}, as well as by only listing one example of each gauge equivalent class of PSGs.  The tables are separated by the most relevant limits of the stuffed honeycomb lattice. Table \ref{QSL1} contains all PSGs that allow both nearest neighbor {\it ansatz}, $u_1$ and $u'$; this table contains all triangular lattice PSGs. Tables \ref{QSL2} and \ref{QSL3} list all PSGs without $u'$ or $u_1$, most relevant for the honeycomb and dice limits, respectively.

Note that, in the interest of space, the tables show only the $\epsilon_2$, $g_{\sigma,s}$ and $g_{R,s}$ for each PSG, which are enough to uniquely specify it.  Each PSG also has a set of signs $\{\epsilon_R,\epsilon_{\sigma}, \epsilon_{R\sigma}\, \epsilon_{TR},\epsilon_{T\sigma}\}$ and matrices, $g_{\mathcal{T},s}$.  While gauge equivalent PSGs may look different initially, they all have the same single spinon dispersion and gauge fluxes.  As an example, consider PSG 1 in table \ref{QSL1} and use the uniform gauge transformation, $g = (i\tau_3,i\tau_2,i\tau_2)$. The transformed PSG has $(g_{\sigma,C},g_{\sigma,B},g_{\sigma,A}, g_{R,C}, g_{R}) = (i\tau_2,i\tau_2,-i\tau_2,i\tau_2,i\tau_2)$, and the {\it ansatz} changes to have $u_1$ proportional to $\tau_3$;
however, the single spinon dispersion is completely unchanged.

\subsection{SU(2) gauge flux}\label{su2}

The SU(2) gauge structure of the mean-field theory means that $\mathds{Z}_2$ spin liquids generically have SU(2) gauge fluxes\cite{wen02,lee06} through some plaquettes.  These fluxes provide another tool to characterize spin liquids, and can, unlike the {\it ansatz} \{$u_{ij}$\}, provide a gauge invariant diagnostic.

These gauge fluxes are most straightforwardly understood in the simpler, U(1) spin liquid case.  In U(1) spin liquids, there exists a gauge in which the {\it ansatz} can be written using only complex hopping terms, without any pairing.  The phase of the hopping enters exactly as the electromagnetic gauge field does for electrons via the Peierls substitution.  The flux, then, is the circulation of the gauge field around a closed path, or Wilson loop, found by multiplying the hopping terms around the path and taking the total phase of the loop.  For U(1) spin liquids, this phase is clearly a number.  However, for $\mathds{Z}_2$ spin liquids, we must keep the full SU(2) structure of the {\it ansatze}, and the SU(2) gauge flux is found similarly by multiplying the {\it ansatz} matrices around a loop.  Now, however the flux has both an angle, corresponding to the U(1) phase, and a direction in SU(2) space.

\begin{figure}[htbp]
\begin{center}
\includegraphics[scale = 0.2]{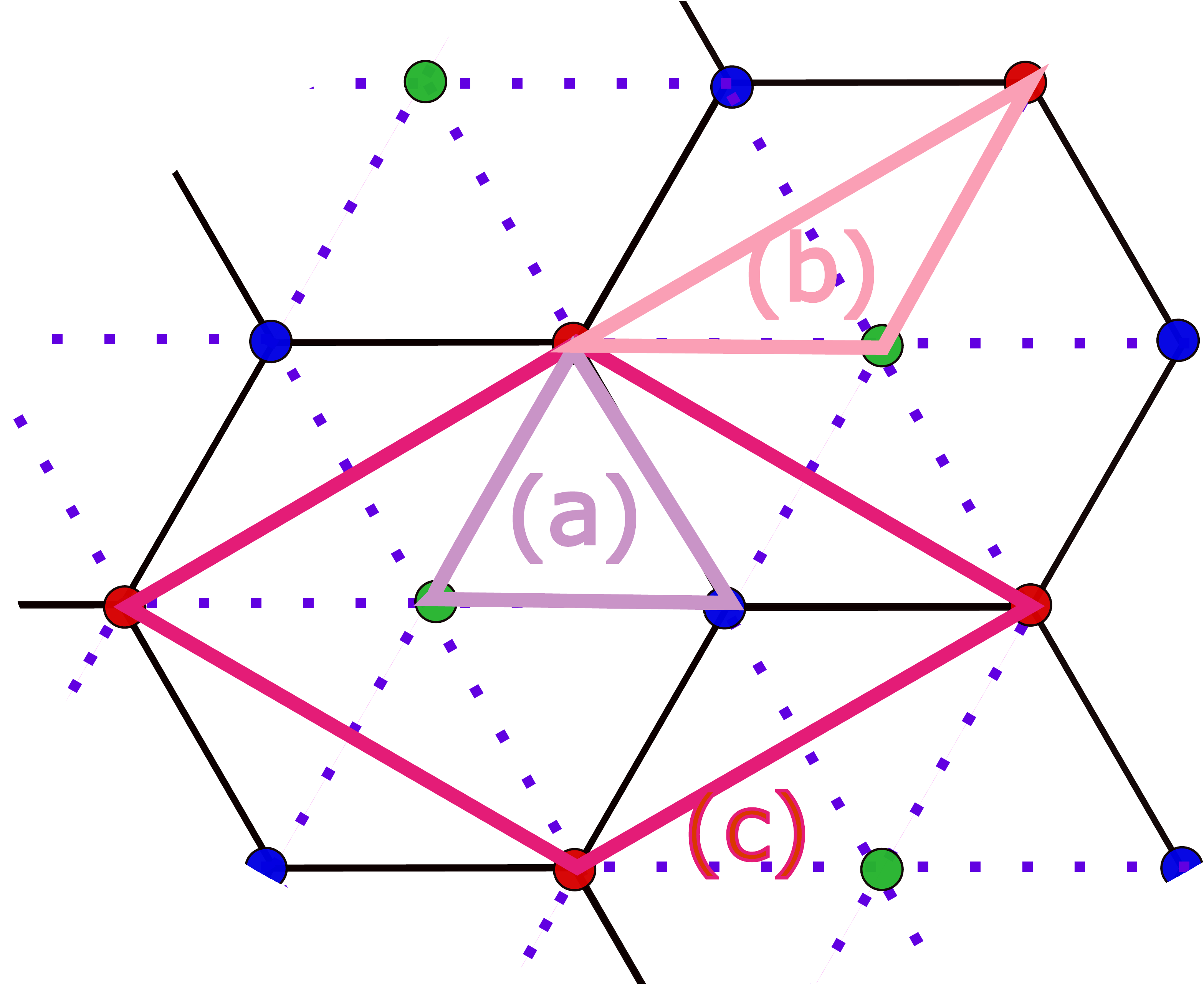}
\end{center}
\caption{\label{fluxfig} The SU(2) gauge flux was calculated through the three loops shown here.  (a) indicates a triangular nearest- neighbor loop, while (b) involves one next-nearest neighbor bond [there are two types of (b) loops, depending on whether the NNN bond is AA/BB or CC], and (c) is a four-site loop involving only next-nearest neighbors [again with two types].}
\end{figure}


The SU(2) gauge flux\cite{bieri16,wen02,lee06} for a Wilson loop starting and ending at a lattice site `i' is defined by multiplying the {\it ansatz} over all the links in the loop,
\beq\label{flux}
F_i = u_{ii_1}u_{i_1i_2}u_{i_2i_3}....u_{i_ni}
\eeq
This definition depends on the gauge of the base site `i', and may be rotated by a local SU(2) gauge transformation. Loops that share a base site may still be directly compared\cite{wen02}.  We consider both even and odd loops, which take the forms\cite{bieri16},
\begin{equation}\label{flux2}
\begin{aligned}
&F_{even} = \rho(\cos\theta + i(\vect{n}\cdot\vect{\sigma})\sin\theta)\\
&F_{odd} =  \rho((\vect{n}\cdot\vect{\sigma})\cos\theta + i\sin\theta),
\end{aligned}
\end{equation}
where `$\rho$' is a scale factor unimportant for our discussion. These quantities are gauge dependent, but their traces are gauge independent and can differentiate the PSGs. 

The SU(2) fluxes are constrained by time-reversal; as $u_{ij}$ changes sign under $\mathcal{T}$, $F_{even}$ is left invariant, while $F_{odd}$ changes sign.  The even loops are therefore constrained to have a director ${\bf n}$ parallel to $g_{\mathcal{T},s}$ on the base site, while odd loops must have the director perpendicular to $g_{\mathcal{T},s}$ and the flux angle, $\theta = 0$ or $\pi$.

\begin{figure}[htbp]
\begin{center}
\includegraphics[scale = 0.2]{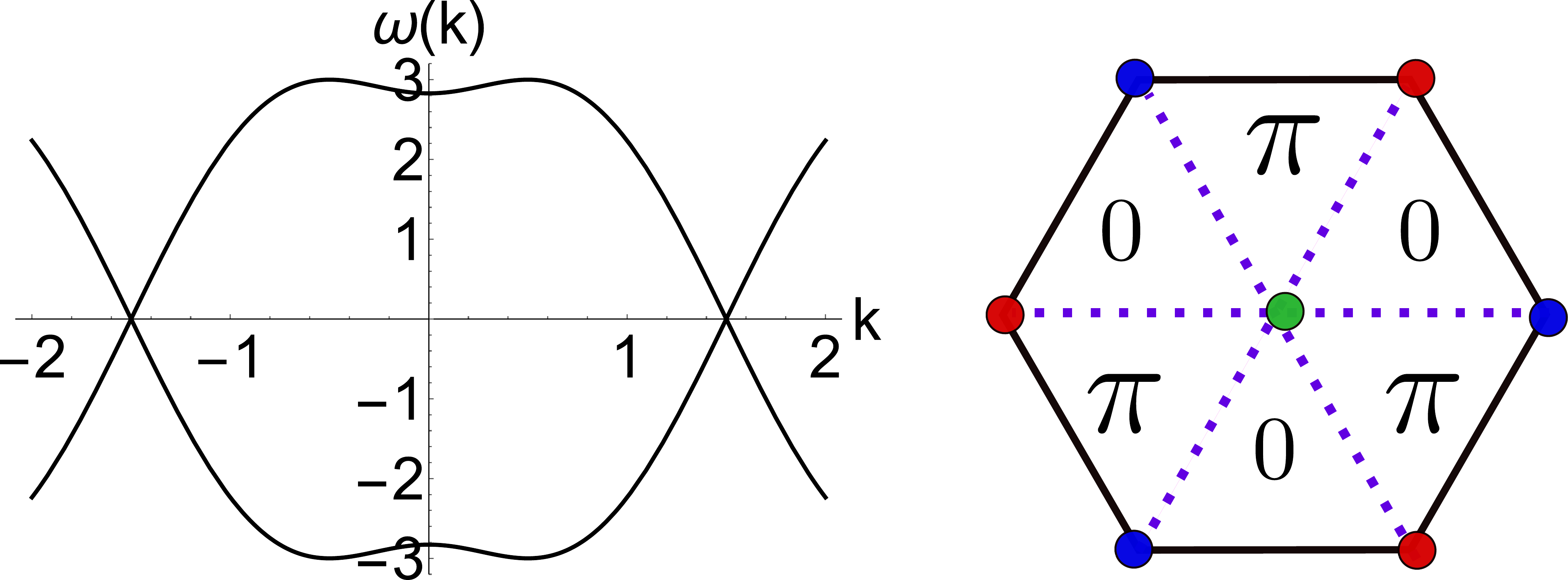}
\end{center}
\caption{\label{fluxdsl} The Dirac spin liquid on the triangular lattice is found as a limiting case of several stuffed honeycomb spin liquids, where only the nearest neighbor $u_1 = u'$'s remain.  (Left) The single spinon dispersion along $(k,0)$, showing the Dirac cones. (Right) The staggered $\pi$ flux structure.}
\end{figure}

There are three relevant plaquettes that we considered, indicated in Fig.\ref{fluxfig}, with both nearest- and next-nearest-neighbor links included. They all have the same base point, which is essential to compare the fluxes. The fluxes through the up and down nearest-neighbor triangles (a) are indicated for the PSGs in Table \ref{QSL1}, where if they have the same directors, the fluxes are either $(0,0)$ or $(\pi,0)$.  If the fluxes have noncollinear directors, then they are labeled `NC'.  Note that if \emph{all} directors are collinear, one can always choose mean field {\it ansatze} $u_{ij}$ (through appropriate SU(2)gauge transformations) of the form $ie^{i\phi_{ij}\tau_3}$. These are now clearly invariant under a global U(1) transformation, and thus are U(1) spin liquids.  The $(\pi,0)$ staggered flux structure is shared by the U(1) Dirac spin liquid (DSL) found on the triangular lattice \cite{lu16,wietek17,iqbal16,shijie19}, as shown in Fig. \ref{fluxdsl}.  The first three PSGs in Table \ref{QSL1} all share this staggered nearest-neighbor flux structure, and in fact are all descendants of the triangular lattice DSL.  However, plaquettes including next-nearest-neighbor links introduce noncollinear fluxes, which breaks the gauge structure down to $\mathds{Z}_2$.


\subsection{Connections to previous work}

The stuffed honeycomb lattice contains the triangular, honeycomb and dice lattices as special limits.  The triangular lattice has a higher symmetry, which requires enforcing $u_1 = u'$, $\lambda_B = \lambda_C$ and $u_2 = u_C$, while the honeycomb lattice has no links to the C spins, $\lambda_C = u' = u_C = 0$, which therefore form a flat band coexisting with the AB spin liquid. Both the triangular and honeycomb lattices have previously been treated with PSG analysis, and we can identify the stuffed honeycomb PSGs with their previously examined limiting cases.
The $J_1-J_2$ triangular lattice is expected to have a Dirac spin liquid for intermediate $J_2/J_1$\cite{wietek17,iqbal16,shijie19}, which we will discuss in detail in the next section, as this spin liquid is the limiting case for PSGs 1-3 in Table \ref{QSL1}.  Adding a ring exchange term favors a spinon Fermi surface, which was found to be the uniform resonating valence bond (RVB) state with uniform real hopping \cite{motrunich05, lu16, bieri16, wen18} that corresponds to our PSG 4 and 6 in the triangular limit; there is also a small region of parameter space in which VMC calculations find a $d+id$ quadratic band touching spin liquid\cite{mishmash13, bieri16} that corresponds to our PSG 5 in the triangular limit, with parabolic spinon bands touching at the $\Gamma$ point.  Finally, an f-wave state was found for ring exchange with ferromagnetic $J_1$ \cite{grover10}, which corresponds to PSG 2, in the triangular limit.

The $J_1-J_2$ honeycomb lattice may host a sublattice pairing state (SPS) spin liquid \cite{lu11}, as found for intermediate $J_2/J_1$ in variational Monte Carlo\cite{clark11}. We find that the gapped SPS on the honeycomb lattice is in fact smoothly connected to the uniform RVB spinon Fermi surface on the triangular lattice limit, as both are limits of PSG 4.  In the honeycomb limit, PSG 4 corresponds to the PSG for the SPS \cite{lu11}, with a flat C band. This correspondence is not immediately obvious, but the two can be related by first doing a uniform gauge rotation about the $\tau_1$ direction ($g = \exp[i \pi \tau_1/4]$) on our SPS {\it ansatz} followed by a second transformation with $g = (i\tau_1,i\tau_3,i\tau_1)$.  After these transformations, we recover the SPS ansatz exactly as found by Lu and Ran\cite{lu11}.  Alternately, a $d\pm id$ state was recently proposed on the honeycomb lattice \cite{ferrari17}, which corresponds to the honeycomb limit of PSGs 5 and 7.

\begin{figure}[t]
\captionsetup[subfigure]{justification=centering}
\centering
\includegraphics[scale=0.1]{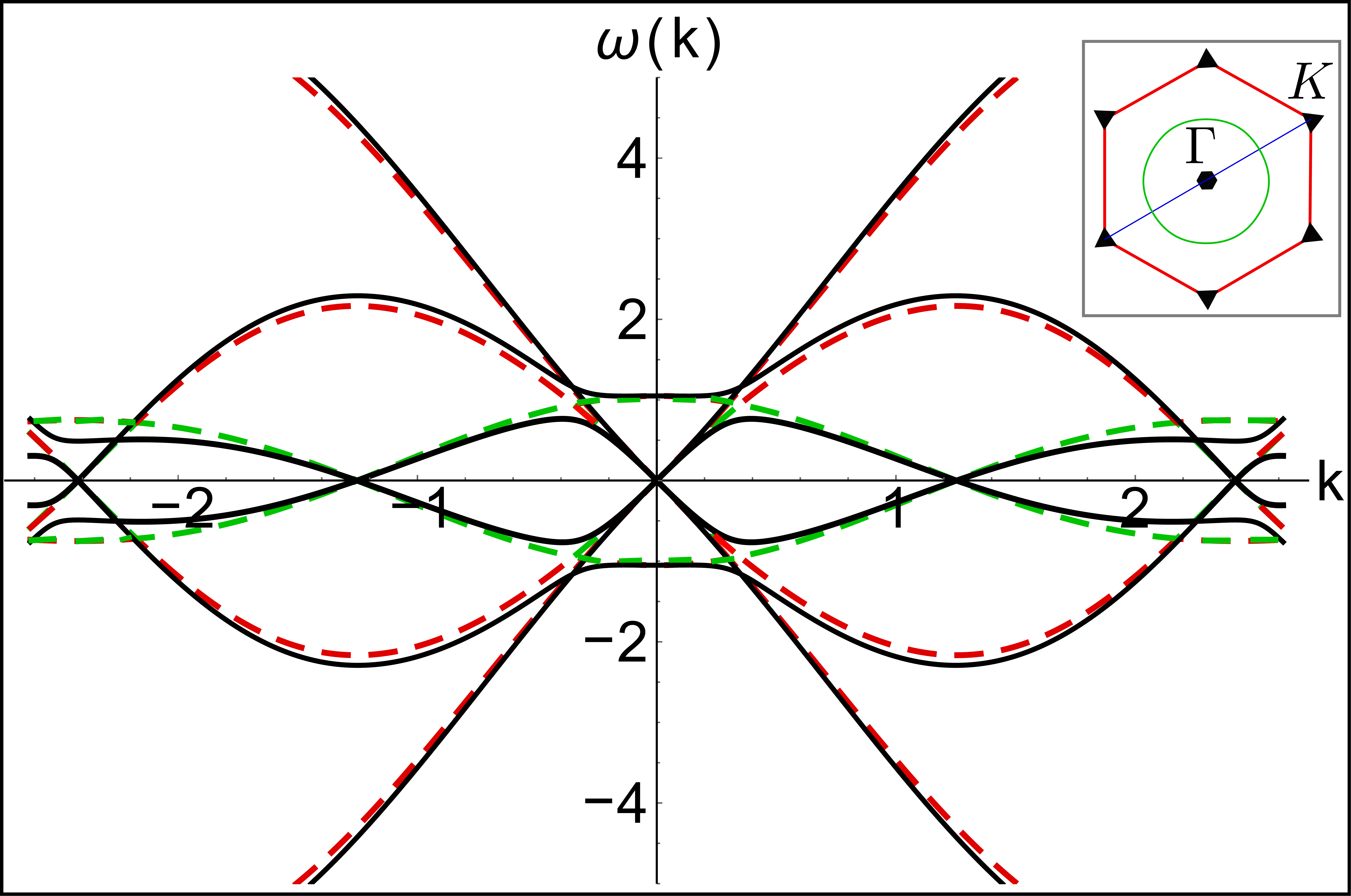}
\caption{Single spinon dispersion for PSG 7, which is competitive in the honeycomb limit \cite{ferrari17}, and may capture the physics of LiZn$_2$Mo$_3$O$_8$.  Here, we consider a small $u'$ that hybridizes the otherwise decoupled AB (red) and C (green) sublattices. For $u' = 0$, the AB spins form a Dirac ($d+id$) spin liquid with doubly degenerate Dirac cones at the $\Gamma$ point and single Dirac cones at the $K$ and $K'$ points, as indicated in the inset; the overall bandwidth is governed by $u_1$. The decoupled C spins form a uniform RVB state with a spinon Fermi surface occupying half the Brillouin zone, whose bandwidth is governed by $u_C$.  As the low energy spinons are separated in momentum space, the effect of even fairly large $u' = .4$ (black) does not lead to significant hybridization. Therefore, it is plausible that the two spin liquids remain relatively decoupled, even out to large $J'$.  There are three relevant temperature scales.  For temperatures greater than the AB bandwidth, all spins are essentially free.  Between this energy scale and the C spin bandwidth, the AB spins will form a correlated, gapless spin liquid, while the C spins remain effectively free.  Finally, at low temperatures, the C spins also form a gapless quantum spin liquid that remains mostly decoupled from the AB spin liquid.  The other parameters used to plot this dispersion are $\lambda_C = -0.15, u_1 = 2$, $u_2 = 0.5$, $u_C = 0.2$. (Inset) The Brillouin zone for the 3-site unit cell with the location of the Dirac nodes and the spinon Fermi surface explicitly shown. The main figure plots the dispersion along the blue line.}
\label{dslpsg7}
\end{figure}

One interesting result here is that there is no PSG known to be competitive in both the $J_1-J_2$ triangular and honeycomb lattices, from which we can conclude that there \emph{cannot} be a single spin liquid connecting the two limits of the $J_1-J'-J_2$ stuffed honeycomb lattice.  The classical phase diagram is quite complicated in between the two limits, with a number of noncollinear and noncoplanar classical phases that are not likely to survive quantum fluctuations\cite{js18}; however, even if there is a continuous region of quantum disorder, there must be a topological phase transition between the two spin liquids.

Another interesting connection is to the cluster magnet LiZn$_2$Mo$_3$O$_8$ \cite{sheckelton12,mourigal14,sheckelton14}, which may realize this $J_1-J'-J_2$ stuffed honeycomb lattice close to the honeycomb limit\cite{flint13}. In LiZn$_2$Mo$_3$O$_8$, the AB spins vanish below 100K and are not even seen in neutron scattering \cite{mourigal14}, while the C spins appear as free spins in the intermediate temperature susceptibility, and are quantum disordered at low T\cite{mourigal14}. One possible explanation is that there is a single spin liquid with two energy scales: a large AB bandwidth and a small C bandwidth, with relatively weak hybridization governed by $u'$.  All three spin liquids with competitive honeycomb energies, PSGs 4,5 and 7, capture slightly different versions of this basic picture.  The AB spins in PSG 4 are gapped, in the SPS state, while the AB spins of PSGs 5 and 7 form Dirac cones.  The C spins of PSGs 4 and 7 form the uniform RVB state (PSG 4 in the triangular limit) out of just $u_C$, which has a spinon Fermi surface filling half the Brillouin zone, while PSG 5 has C spins that form the quadratic band touching spin liquid (still PSG 5) with just $u_C$.  For PSGs 4 and 7, low energy AB and C spinons do not coexist in momentum space, and so even relatively large $J'$ is expected to leave the two sets of sublattices relatively decoupled.  In this case, for intermediate temperatures, the AB spins form a correlated spin liquid, while the C spins behave like free spins.  The spinon dispersion for PSG 7 is shown in Fig. \ref{dslpsg7}, where this separation of energy scales and weak hybridization is apparent.  The story for PSG 5 is more complicated, as the Dirac points of the AB and C spins coexist and are partially gapped out by the $u'$ hybridization. 

\subsection{Triangular lattice descendants}\label{Tlim}

The nature of the spin liquid on the $J_1-J_2$ triangular lattice is widely debated\cite{kaneko14,shimada18,saadatmand15,itou08}, but many numerical methods have found the U(1) DSL\cite{wietek17,iqbal16,shijie19}, shown in Fig.\ref{fluxdsl}. Three of the spin liquids on the stuffed honeycomb lattice reduce to the triangular lattice DSL in the triangular limit; these are the first three PSGs in table \ref{QSL1}, which we will discuss in detail in this section.

\begin{figure}[htbp]
\begin{center}
\includegraphics[scale = 0.17]{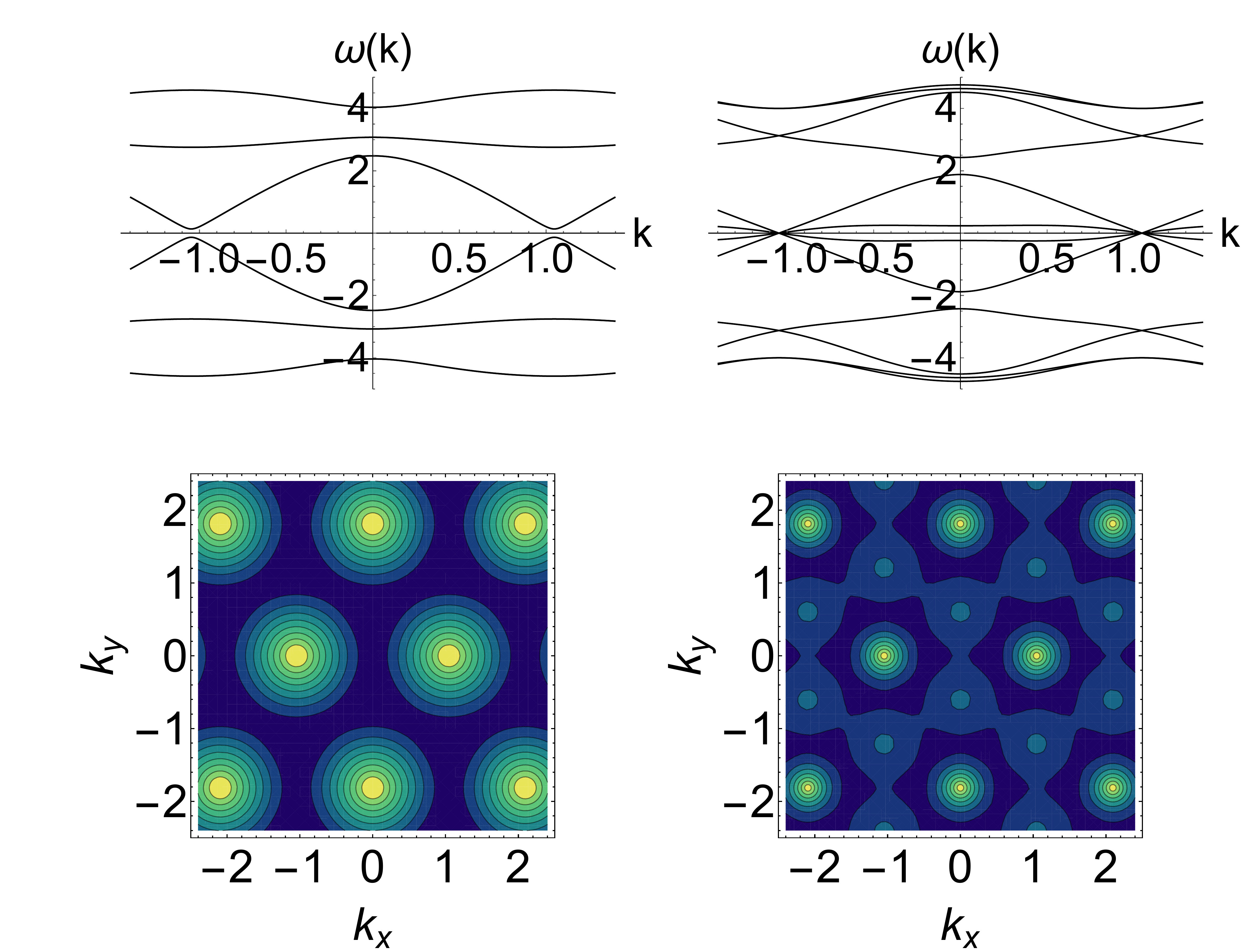}
\end{center}
\caption{\label{psg12} Single spinon spectra for two stuffed honeycomb spin liquids that can arise from the triangular DSL: (left) PSG 1 and (right) PSG 2.   (Top) Spinon bands along the $(k,0)$ axis. (Bottom) Contour plots of the lowest lying positive energy bands. Bright (yellow) points indicate the Dirac points, where the gap vanishes for PSG 2 and reaches a minimum for PSG 1. The color scale varies from bright (yellow) to dark (blue) as the energy increases, and shows the six-fold symmetry of the Dirac cones.  Both PSGs have a staggered $\pi$- flux structure as shown in Fig. \ref{fluxdsl}.}
\end{figure}

\begin{figure}[htbp]
\begin{center}
\includegraphics[scale = 0.12]{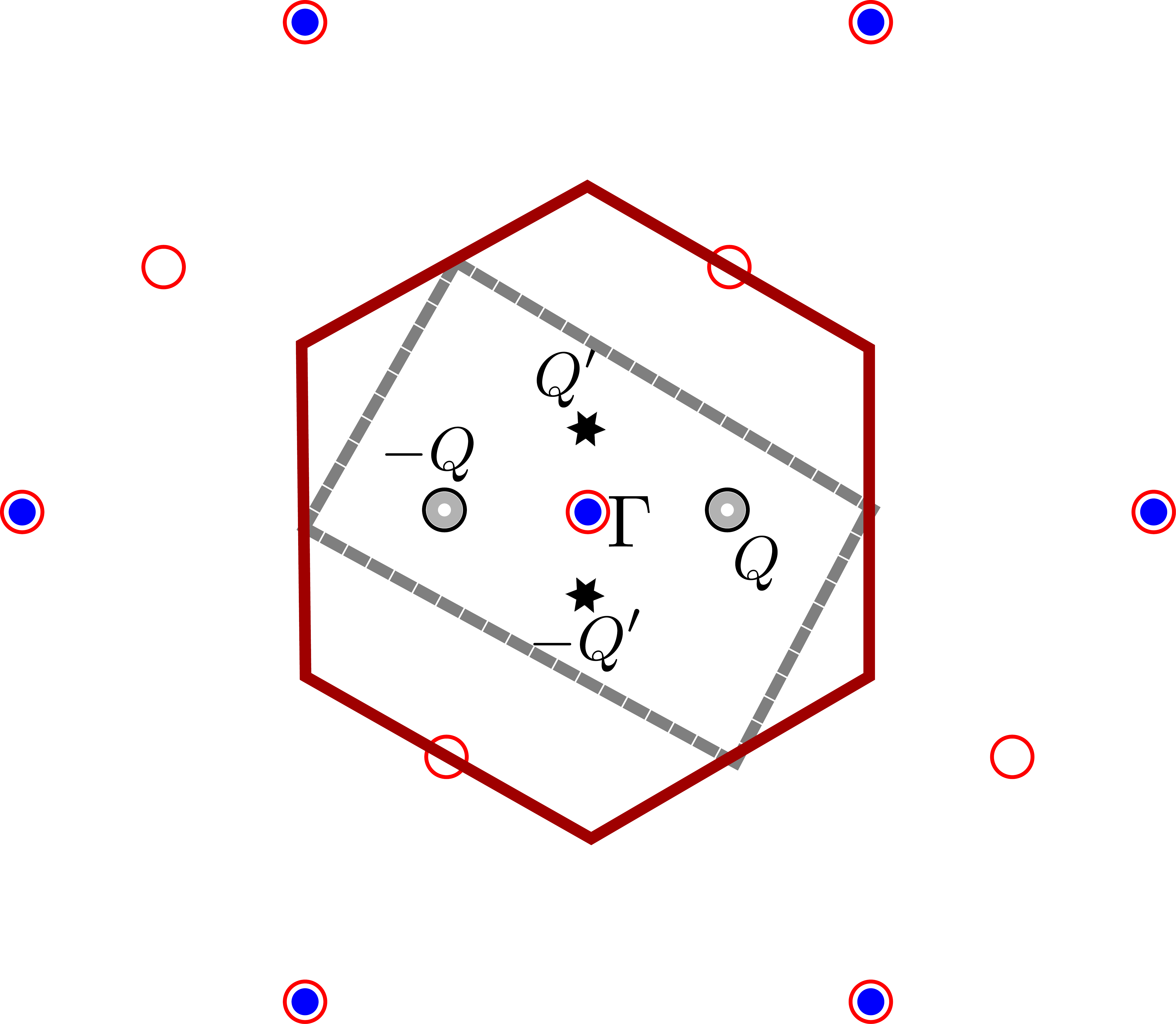}
\end{center}
\caption{\label{BZ} Locations of the Dirac cones for the three descendants of the triangular DSL. The red hexagon shows the original Brillouin zone (BZ) of the stuffed honeycomb lattice, while the grey rectangle is the BZ for the extended six-site unit cell.  Empty (red) circles show the rectangular reciprocal lattice, while filled (blue) disks are the original hexagonal reciprocal lattice.  $\pm Q$ =($\pi$/3,0)] are the locations of the Dirac points for PSG 1 and 2, which are six-fold symmetric. PSG 3 has four Dirac points in the BZ: $\pm Q$ and $\pm Q' = (0,\frac{\pi}{3\sqrt{3}})$. While the Dirac cones at $\pm Q$ are six-fold symmetric and doubly- degenerate, the cones at $\pm Q'$ are three-fold symmetric, and singly degenerate.}
\end{figure}

\begin{figure}[h]
\captionsetup[subfigure]{justification=centering}
\centering
\includegraphics[scale = 0.072]{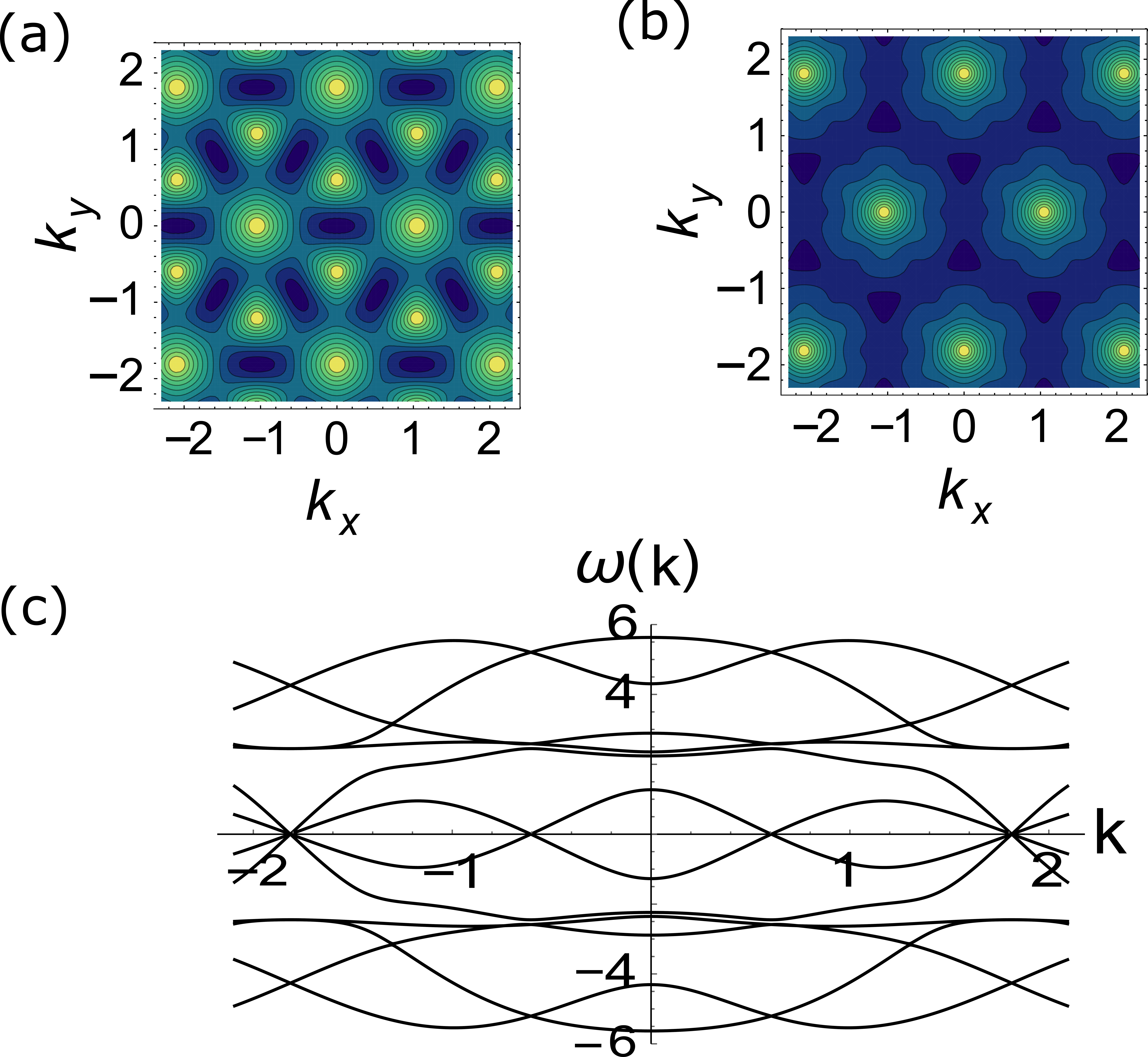}
\caption{Spinon dispersion for PSG 3. (Top) Contour plots for the lowest two positive energy bands, with (a) the lowest and (b) the next lowest. Bright spots indicate the location of the Dirac cones, while the color of the contour changes from bright (yellow) to dark (blue) as the energy increases, showing the six-fold and three-fold symmetries of the Dirac cones.  Note that the six-fold symmetric Dirac cones are doubly degenerate.  (c) Spinon dispersion along the $(0,k)$ axis, for a generic set of parameters, also used in (a) and (b): $u_1 = 1$, $u_2 = -0.45$, $u_C = 1$, $u' = 1.23$.  }
\label{dslpsg3}
\end{figure}

While all three PSGs yield the same U(1) Dirac dispersion in the triangular limit, PSG 1 is generically gapped, while PSGs 2 and 3 preserve the Dirac nodal structure. PSG 1 corresponds to \#20 in table I of  ref. \onlinecite{lu16}, up to a gauge transformation. All three PSGs become $\mathds{Z}_2$ spin liquids when any next-nearest-neighbor {\it ansatz} are allowed. The single spinon dispersions for PSGs 1 and 2 are shown in Fig. \ref{psg12} and in Fig. \ref{dslpsg3} for PSG 3.  Both PSG 1 and 2 have two (possibly gapped) Dirac points in the rectangular Brillouin zone.  Each Dirac point is doubly degenerate and six-fold symmetric.  For PSG 1, all bands are doubly degenerate. The Dirac cones remain gapless along the line $\lambda_B^2\lambda_C + 3\lambda_Cu_1^2 - 6\lambda_Bu'^2 = 0$, where the $u$'s and $\lambda$'s are the amplitudes of the relevant {\it ansatz} terms. The single spinon dispersions for PSG 2 and 3 explicitly break translation symmetry, as they are not doubly degenerate; the symmetry will be restored in the physical two spinon spectrum. The original and halved Brillouin zones are shown in Fig.\ref{BZ}, along with the Dirac point locations for all three PSGs. Translation invariance can explicitly be restored by a gauge transformation shifting half of the bands, essentially unfolding the bandstructure.

PSG 3 also has a Dirac dispersion, but now generically has four Dirac points in the Brillouin zone, as shown in Fig. \ref{dslpsg3}.  Two of the Dirac cones are six-fold symmetric and doubly degenerate, occurring at the same locations as those in PSG 1 and 2.  However, there are also two new three-fold symmetric, singly degenerate Dirac cones located between the six-fold points.  In the triangular limit, these also become six-fold symmetric and are just shifted copies of the others.  Again, PSG 3 has $\epsilon_2 = -1$, and so the single spinon dispersion is not translation invariant; the translation symmetry can similarly be restored by an appropriate gauge transformation. 

\section{Conclusions}
\label{concl}

We have enumerated all possible spin liquids on the stuffed honeycomb lattice with SU(2) symmetric interactions, with particular emphasis on the spin liquids relevant near the limiting cases of the honeycomb and triangular lattices. Notably, we find three $\mathds{Z}_2$ descendants of the U(1) DSL on the triangular lattice, and discuss how two potential  spin liquids on the honeycomb lattice may also be relevant for LiZn$_2$Mo$_3$O$_8$. This analysis provides a useful starting point for variational Monte Carlo studies of spin liquid stability, which can treat the {\it ansatz} parameters as variational parameters while enforcing the constraint of half-filling exactly via the Gutzwiller projection.  More generally, understanding what spin liquids are possible is essential to interpreting numerical solutions.  It is an interesting open question how far the U(1) DSL might survive as the system is tuned away from the triangular limit, and if any of these descendants become relevant.

\vskip 0.5in
\noindent{\bf Acknowledgments}
\vskip 0.04in
We are grateful for useful discussions with Bryan Clark and Victor Quito. J.S. and R.F. were supported by NSF DMR-1555163. R.F. also acknowledges the hospitality of the Aspen Center for Physics, supported by National Science Foundation Grant No. PHY-1607611.

\appendix

\section{Derivation of gauge representations}

\subsection{Space group symmetries}
\label{appSG}

In this appendix, we derive the gauge representations for the space group elements of the stuffed honeycomb lattice. The representations for the two translation operators was shown in the main text.  Here, we explicitly incorporate the algebraic relations \eqref{alg1} and \eqref{alg2} to find the choices for the rotation and reflection operations.

To do so, we must know how the space group elements transform the spatial coordinates, $(x,y,s)$:
\begin{equation}
\begin{aligned}
&C_6/C_6^{-1}:  (x,y,C) \rightarrow
\begin{cases}
    (-y,x+y,C),& \text{$C_6$}\\
    (x+y,-x,C),              & \text{$C_6^{-1}$}
\end{cases}\\
&\,\,\,\,\,\,\,\,\,\,\,\,\,\,\,\,\,\,:(x,y,B)\rightarrow
\begin{cases}
    (-y-1,x+y+1,A),& \text{$C_6$}\\
    (x+y,-x,A),              & \text{$C_6^{-1}$}
\end{cases}\\
&\,\,\,\,\,\,\,\,\,\,\,\,\,\,\,\,\,\,:(x,y,A)\rightarrow
\begin{cases}
    (-y,x+y,B),& \text{$C_6$}\\
    (x+y,-x-1,B),              & \text{$C_6^{-1}$}
\end{cases}\\
&\sigma : (x,y,C)\rightarrow (y,x,C)\\
&\,\,\,\,: (x,y,B)\rightarrow (y,x,B)\\
&\,\,\,\,: (x,y,A)\rightarrow (y-1,x+1,A)\\
&T_{\hat{x}} : (x,y,s) \rightarrow (x+1,y,s)\\
&T_{\hat{y}} : (x,y,s) \rightarrow (x,y+1,s).
\end{aligned}
\end{equation}
If we consider a particular algebraic relation, we can use eq. \eqref{psg_mult} to find the conditions on the gauge elements.  Eq. \eqref{rel1} was used to fix the translation gauge elements. Here, we use the relations in eqs.\eqref{rel2}  and \eqref{rel5} to obtain the spatial dependence of the mirror plane gauge element, $g_{\sigma}(x,y,s)$. Beginning with eq. \eqref{rel2}, we have
\begin{equation}\label{R1}
\begin{split}
g_{\sigma}(x,y,s)
\left[g_{\hat{x}}(y,x,s)(\delta_{s,C}+\delta_{s,B})\right.\\\left.+g_{\hat{x}}(y-1,x+1,s)\delta_{s,A}\right] \\= \epsilon_{\sigma 2}g_{\hat{y}}(x,y,s)g_{\sigma}(x,y-1,s).
\end{split}
\end{equation}
Using our expressions for $g_{\hat x}$ and $g_{\hat y}$, we find
\begin{equation}\label{R2}
\begin{split}
g_{\sigma}(x,y,s) = \epsilon_{\sigma 2}\epsilon_2^xg_{\sigma}(x,y-1,s)\\
\implies  g_{\sigma}(x,y,s) = \epsilon_{\sigma 2}^y\epsilon_2^{xy}g_{\sigma}(x,s),
\end{split}
\end{equation}
where $g_\sigma(x,s)=g_\sigma(x,y=0,s)$.  To further constrain $g_\sigma$, we use eq. \eqref{rel5} and proceed similarly.  We treat each sublattice independently.
For s=A,
\begin{equation}\label{s3}
\begin{split}
&g_{\sigma}(x,y,A)g_{\sigma}(y-1,x+1,A) = \epsilon_{\sigma}\mathds{1}_2\\
\implies& \epsilon_{\sigma 2}^y \epsilon_2^{xy} \epsilon_{\sigma 2}^{x+1} \epsilon_2^{(x+1)(y-1)} g_{\sigma}(x,A)g_{\sigma}(y-1,A) = \epsilon_{\sigma}\mathds{1}_2.
\end{split}
\end{equation}
For s=B,
\begin{equation}\label{s2}
\begin{split}
&g_{\sigma}(x,y,B)g_{\sigma}(y,x,B) = \epsilon_{\sigma}\mathds{1}_2\\
\implies& \epsilon_{\sigma 2}^y \epsilon_2^{xy} \epsilon_{\sigma 2}^x \epsilon_2^{xy} g_{\sigma}(x,B)g_{\sigma}(y,B) = \epsilon_{\sigma}\mathds{1}_2.
\end{split}
\end{equation}
For s=C,
\begin{equation}\label{s1}
\begin{split}
&g_{\sigma}(x,y,C)g_{\sigma}(y,x,C) = \epsilon_{\sigma}\mathds{1}_2\\
\implies& \epsilon_{\sigma 2}^y \epsilon_2^{xy} \epsilon_{\sigma 2}^x \epsilon_2^{xy} g_{\sigma}(x,C)g_{\sigma}(y,C) = \epsilon_{\sigma}\mathds{1}_2.
\end{split}
\end{equation}

Equations \eqref{s1},\eqref{s2} and \eqref{s3} can be satisfied by,
\begin{eqnarray}
\begin{aligned}
g_{\sigma}(x,C) = \epsilon_{\sigma 2}^x g_{\sigma,C}\\
g_{\sigma}(x,B) = \epsilon_{\sigma 2}^x g_{\sigma,B}\\
g_{\sigma}(x,A) =\epsilon_2^x\epsilon_{\sigma 2}^x g_{\sigma,A}.
\end{aligned}
\end{eqnarray}

The sign $\epsilon_{\sigma 2}$ can be eliminated via the staggered gauge transformation, $g(x,y,s) = (-1)^x$; note that this transformation does not affect the previously determined gauge representations by more than an overall sign.
Finally, we have the following PSG representations for $g_{\sigma}$:
\begin{equation}
\begin{aligned}
&g_{\sigma}(x,y,C) = \epsilon_2^{xy}g_{\sigma,C}\\
&g_{\sigma}(x,y,B) = \epsilon_2^{xy}g_{\sigma,B}\\
&g_{\sigma}(x,y,A) = \epsilon_2^{x(y+1)}g_{\sigma,A}
\end{aligned}
\end{equation}

Eq.\eqref{rel3} allows us to fix the gauge element associated with the rotation operator $C_6$. With the gauge representations for translation operators already established in eq.\eqref{T}, the left side of eq. \eqref{rel3} becomes,
\begin{equation}\label{TR-i}
\begin{split}
Q_{\hat{y}}Q_{C_6}& = (g_{\hat{y}}T_{\hat{y}}g_{C_6}T_{\hat{y}}^{-1}, T_{\hat{y}}C_6)\\
& =  \omega(\hat{y},C_6) \omega^{-1}(C_6,\hat{x})Q_{C_6}Q_{\hat{x}}\\
& = \epsilon_{R1}Q_{C_6}Q_{\hat{x}}
\end{split}
\end{equation}

The first relation in \eqref{rel3} can then be expanded as,
\begin{equation}\label{TR-ii}
\begin{split}
g_{\hat{y}}(x,y,&s)g_{C_6}(x,y-1,s)\\&= \epsilon_{R1}g_{C_6}(x,y,s)[g_{\hat{x}}(x+y,-x,s)\delta_{s,C}\\  &\quad+ g_{\hat{x}}(x+y,-x,s+1)\delta_{s,B} \\ &\quad+ g_{\hat{x}}(x+y,-x-1,s-1)\delta_{s,A}]\\
 \implies & (\epsilon_{2})^x g_{C_6}(x,y-1,s) = \epsilon_{R1}g_{C_6}(x,y,s)\\
 \implies & g_{C_6}(x,y,s) = (\epsilon_{R1})^y(\epsilon_2)^{xy} g_{C_6}(x,s).
\end{split}
\end{equation}

Now using the second equality in relation \eqref{rel3},
\begin{equation}\label{TR-iii}
Q_{C_6}Q_{\hat{x}} = \omega(C_6,\hat{x})Q_{T_{\hat{x}} C_6 T_{\hat{y}}} = \epsilon_{R2}Q_{T_{\hat{x}}}Q_{C_6}Q_{T_{\hat{y}}},
\end{equation}
which explicitly becomes,
\begin{equation}\label{TR-iv}
\begin{split}
g_{C_6}(x,y,s)g_{\hat{x}} &= \epsilon_{R2}g_{\hat{x}}T_{\hat{x}}g_{C_6}(x,y,s)T_{\hat{x}}^{-1}\!T_{\hat{x}} C_6g_{\hat{y}}(x)C_6^{-1}\!T_{\hat{x}}^{-1}\\
g_{C_6}(x,y,s)& = \epsilon_{R2}g_{C_6}(x-1,y,s) g_{\hat{y}}(x+y-1)\\
& = \epsilon_{R2} \epsilon_{2}^{x+y-1} g_{C_6}(x-1,y,s)
\end{split}
\end{equation}

We can then find the x-dependence as,
\begin{equation}\label{TR-v}
\begin{split}
g_{C_6}(x,y,s) & =\epsilon_{R2}(\epsilon_{2})^{x+y-1}g_{C_6}(x-1,y,s)\\
& = (\epsilon_{R2})^2(\epsilon_{2})^{2x+2y-3}g_{C_6}(x-2,y,s)\\
&\,\,\,\,\,\, \vdots\\
& = (\epsilon_{R2})^n(\epsilon_{2})^{nx+ny-\sum_{k=1}^n k}g_{C_6}(x-n,y,s)\\
& = (\epsilon_{R2})^x(\epsilon_{2})^{xy + x(x-1)/2}g_{C_6}(y,s).
\end{split}
\end{equation}

$g_{C_6}$ must satisfy eq.\eqref{TR-ii} and \eqref{TR-iv} simultaneously, and hence takes the form,
\begin{equation}\label{TR-vi}
g_{C_6}(x,y,s) = (\epsilon_{R1})^y(\epsilon_{R2})^x(\epsilon_{2})^{xy + x(x-1)/2}g_{R,s}.\\
\end{equation}
We can again remove the sign $\epsilon_{R2}$ using the staggered gauge transformation $g(x,y,s) = (-1)^{x+y}$, which again leaves all previously determined gauge representations unmodified up to a sign.  We then have,
\begin{equation}\label{TR-vii}
g_{C_6}(x,y,s) =  (\epsilon_{R1})^y(\epsilon_{2})^{xy + x(x-1)/2}g_{R,s}.
\end{equation}

We can also find a representation for the symmetry operator $R\sigma$ to explicitly satisfy Eq.\eqref{rel8}. We compute this representation for each individual sublattice by using the group multiplication defined in Eq.\eqref{psg_mult}. For s=A,
\begin{equation}\label{RR3}
\begin{split}
g_{R\sigma}(x,y,A) &= g_{C_6}(x,y,A)g_{\sigma}(x+y,-x-1,B)\\
& = \epsilon_{R1}^y\epsilon_2^{xy+x(x-1)/2 + (-x-1)(x+y)}\epsilon_2^{xy}g_{R,A}g_{\sigma,B}\\
& = \epsilon_{R1}^y\epsilon_2^{-x(3+x)/2 - y}g_{R,A}g_{\sigma,B}.
\end{split}
\end{equation}
\noindent For s=B,
\begin{equation}\label{RR2}
\begin{split}
g_{R\sigma}(x,y,B) &= g_{C_6}(x,y,B)g_{\sigma}(x+y,-x,A)\\
& = \epsilon_{R1}^y\epsilon_2^{xy+x(x-1)/2 + (-x+1)(x+y)}\epsilon_2^{xy}g_{R,B}g_{\sigma,A}\\
& = \epsilon_{R1}^y\epsilon_2^{x(1-x)/2 + y}g_{R,B}g_{\sigma,A}.
\end{split}
\end{equation}
\noindent Finally, for s=C,
\begin{equation}\label{RR1}
\begin{split}
g_{R\sigma}(x,y,C) &= g_{C_6}(x,y,C)g_{\sigma}(x+y,-x,C)\\
& = \epsilon_{R1}^y\epsilon_2^{xy+x(x-1)/2 - x(x+y)}g_{R,C}g_{\sigma,C}\\
& = \epsilon_{R1}^y\epsilon_2^{x(x+1)/2}g_{R,C}g_{\sigma,C}.
\end{split}
\end{equation}
The above expressions for $g_{R\sigma}$ are constrained by eq.\eqref{rel8}, which can also eliminate $\epsilon_{R1}$, as we now show. For s=C,
\begin{equation}\label{RR-ii}
\begin{split}
&g_{R\sigma}(x,y,C)(R\sigma)g_{R\sigma}(x,y,C)(R\sigma)^{-1}  = \pm \mathds{1}_2\\
& g_{R\sigma}(x,y,C)g_{R\sigma}(-x,x+y,C) = \pm \mathds{1}_2\\
\implies & \epsilon_{R1}^{x}\epsilon_2^{x}(g_{R,C}g_{\sigma,C})^2 = \pm\mathds{1}_2
\end{split}
\end{equation}
This equation forces $\epsilon_{R1} = \epsilon_2$.

For s=A,
\begin{equation}\label{RR-iii}
\begin{split}
&g_{R\sigma}(x,y,A)(R\sigma)g_{R\sigma}(x,y,B)(R\sigma)^{-1}  = \pm \mathds{1}_2\\
& g_{R\sigma}(x,y,A)g_{R\sigma}(-x-1,x+y+1,B) = \pm \mathds{1}_2\\
& \epsilon_{R1}^{x+1}\epsilon_2^{x}(g_{R,A}g_{\sigma,A}g_{R,B}g_{\sigma,B}) = \pm\mathds{1}_2\\
\implies & g_{R,A}g_{\sigma,B}g_{R,B}g_{\sigma,A} = \epsilon_2\epsilon_{R\sigma}\mathds{1}_2.
\end{split}
\end{equation}
s=B follows similar steps as above with the sublattice index A and B swapped in eq. \eqref{RR-iii}. The use of eq.\eqref{gR} changes the gauge such that $g_{R,A}$ = $g_{R,B} = g_R$. So we can simply replace both $g_{R,B}$ and $g_{R,A}$ by $g_{R}$. Eq. \eqref{rel6} is trivially satisfied and does not impose any further constraints.


The gauge representations of all of the space group symmetry operators is now,
\begingroup
\allowdisplaybreaks
\begin{align*}
g_{\hat{x}}(x,y,s) =  \mathds{1}_2 \\
g_{\hat{y}}(x,y,s) = (\epsilon_{2})^x \mathds{1}_2\\
g_\sigma(x,y,C) = (\epsilon_{2})^{xy} g_{\sigma,C}\\
g_\sigma(x,y,B) = (\epsilon_2)^{xy} g_{\sigma,B}\\\numberthis
g_\sigma(x,y,A) = (\epsilon_2)^{x(y+1)} g_{\sigma,A}\\
g_{C_6}(x,y,C) =  (\epsilon_{2})^{(x+1)y + x(x-1)/2}g_{R,C}\\
g_{C_6}(x,y,B) =  (\epsilon_{2})^{(x+1)y + x(x-1)/2}g_{R}\\
g_{C_6}(x,y,A) =  (\epsilon_{2})^{(x+1)y + x(x-1)/2}g_{R},
\end{align*}
\endgroup

where $\epsilon_2$ is the only remaining sign, and $g_{\sigma,s}$, $g_R$, $g_{R,C}$ are SU(2) matrices that must satisfy,

\begingroup
\allowdisplaybreaks
\begin{align}\label{PSG}
\begin{split}
(g_{\sigma,s})^2 = \epsilon_{\sigma}\mathds{1}_2\\
g_{R,C}^6 = g_{R}^6 = \epsilon_{R}\mathds{1}_2\\
(g_{R,C}g_{\sigma,C})^2 = \epsilon_{R\sigma}\mathds{1}_2\\
g_Rg_{\sigma,B}g_Rg_{\sigma,A} = g_Rg_{\sigma,A}g_Rg_{\sigma,B} =\epsilon_2\epsilon_{R\sigma}\mathds{1}_2.
\end{split}
\end{align}
\endgroup

\subsection{Time reversal symmetry}
\label{appTR}

Time-reversal acts trivially on the real space lattice, and thus commutes with all space group operations.  Furthermore, once we gauge fix the time-reversal operator to act as in eq.\eqref{Tfunc}, $\mathcal{T}$ acts trivially on any gauge,
\beq
\mathcal{T}g\mathcal{T}^{-1} = (i\tau_2)g^*(-i\tau_2) = g.
\eeq
The commutation relations in eq.\eqref{Tcom} and the above property of time reversal operation yield further constraints on $g_{\mathcal{T}}$ and the resulting signs.  Considering the commutation with the translation operators, we have
\begin{equation}
    \begin{aligned}
    g_{\mathcal{T}}(x,y,s) = \epsilon_{T1}^xg_{\mathcal{T}}(y,s)\\
    g_{\mathcal{T}}(x,y,s) = \epsilon_{T2}^yg_{\mathcal{T}}(x,s),
    \end{aligned}
\end{equation}
which implies that
\begin{equation}\label{Ttrans}
    g_{\mathcal{T}}(x,y,s) = \epsilon_{T1}^x\epsilon_{T2}^yg_{\mathcal{T},s}.
\end{equation}

The commutation relation with $C_6$ gives the following condition,
\begin{equation}\label{Trot}
    \begin{aligned}
    &g_R(x,y,s)\left[g_{\mathcal{T}}(x+y,-x,s)(\delta_{s,C}+\delta_{s,B})\right.\\&\left.+g_{\mathcal{T}}(x+y,-x-1,s)\delta_{s,A}\right] = \epsilon_{TR}g_{\mathcal{T}}(x,y,s)g_R(x,y,s)\\
    \end{aligned}
\end{equation}
Upon substitution of $g_{\mathcal{T}}$ from eq.\eqref{Ttrans}, one can see that this equation is satisfied only for $\epsilon_{T1} = \epsilon_{T2} = 1$ which removes any spatial dependence of $g_{\mathcal{T}}$, although it can still vary between sublattices. The conditions for $g_{\sigma}$ follows straightforwardly from the corresponding commutation relations and the final form of all these resulting constraints is shown in eq.\eqref{TR_Rs}.

\bibliographystyle{apsrev4-1}
\bibliography{stuffedPSG}

\begin{thebibliography}{56}%
\makeatletter
\providecommand \@ifxundefined [1]{%
 \@ifx{#1\undefined}
}%
\providecommand \@ifnum [1]{%
 \ifnum #1\expandafter \@firstoftwo
 \else \expandafter \@secondoftwo
 \fi
}%
\providecommand \@ifx [1]{%
 \ifx #1\expandafter \@firstoftwo
 \else \expandafter \@secondoftwo
 \fi
}%
\providecommand \natexlab [1]{#1}%
\providecommand \enquote  [1]{``#1''}%
\providecommand \bibnamefont  [1]{#1}%
\providecommand \bibfnamefont [1]{#1}%
\providecommand \citenamefont [1]{#1}%
\providecommand \href@noop [0]{\@secondoftwo}%
\providecommand \href [0]{\begingroup \@sanitize@url \@href}%
\providecommand \@href[1]{\@@startlink{#1}\@@href}%
\providecommand \@@href[1]{\endgroup#1\@@endlink}%
\providecommand \@sanitize@url [0]{\catcode `\\12\catcode `\$12\catcode
  `\&12\catcode `\#12\catcode `\^12\catcode `\_12\catcode `\%12\relax}%
\providecommand \@@startlink[1]{}%
\providecommand \@@endlink[0]{}%
\providecommand \url  [0]{\begingroup\@sanitize@url \@url }%
\providecommand \@url [1]{\endgroup\@href {#1}{\urlprefix }}%
\providecommand \urlprefix  [0]{URL }%
\providecommand \Eprint [0]{\href }%
\providecommand \doibase [0]{http://dx.doi.org/}%
\providecommand \selectlanguage [0]{\@gobble}%
\providecommand \bibinfo  [0]{\@secondoftwo}%
\providecommand \bibfield  [0]{\@secondoftwo}%
\providecommand \translation [1]{[#1]}%
\providecommand \BibitemOpen [0]{}%
\providecommand \bibitemStop [0]{}%
\providecommand \bibitemNoStop [0]{.\EOS\space}%
\providecommand \EOS [0]{\spacefactor3000\relax}%
\providecommand \BibitemShut  [1]{\csname bibitem#1\endcsname}%
\let\auto@bib@innerbib\@empty
\bibitem [{\citenamefont {Savary}\ and\ \citenamefont
  {Balents}(2016)}]{savary16}%
  \BibitemOpen
  \bibfield  {author} {\bibinfo {author} {\bibfnamefont {L.}~\bibnamefont
  {Savary}}\ and\ \bibinfo {author} {\bibfnamefont {L.}~\bibnamefont
  {Balents}},\ }\href {\doibase 10.1088/0034-4885/80/1/016502} {\bibfield
  {journal} {\bibinfo  {journal} {Reports on Progress in Physics}\ }\textbf
  {\bibinfo {volume} {80}},\ \bibinfo {pages} {016502} (\bibinfo {year}
  {2016})}\BibitemShut {NoStop}%
\bibitem [{\citenamefont {Yan}\ \emph {et~al.}(2011)\citenamefont {Yan},
  \citenamefont {Huse},\ and\ \citenamefont {White}}]{yan11}%
  \BibitemOpen
  \bibfield  {author} {\bibinfo {author} {\bibfnamefont {S.}~\bibnamefont
  {Yan}}, \bibinfo {author} {\bibfnamefont {D.~A.}\ \bibnamefont {Huse}}, \
  and\ \bibinfo {author} {\bibfnamefont {S.~R.}\ \bibnamefont {White}},\
  }\href@noop {} {\bibfield  {journal} {\bibinfo  {journal} {Science}\ }\textbf
  {\bibinfo {volume} {332}},\ \bibinfo {pages} {1173} (\bibinfo {year}
  {2011})}\BibitemShut {NoStop}%
\bibitem [{\citenamefont {Ran}\ \emph {et~al.}(2007)\citenamefont {Ran},
  \citenamefont {Hermele}, \citenamefont {Lee},\ and\ \citenamefont
  {Wen}}]{ran07}%
  \BibitemOpen
  \bibfield  {author} {\bibinfo {author} {\bibfnamefont {Y.}~\bibnamefont
  {Ran}}, \bibinfo {author} {\bibfnamefont {M.}~\bibnamefont {Hermele}},
  \bibinfo {author} {\bibfnamefont {P.~A.}\ \bibnamefont {Lee}}, \ and\
  \bibinfo {author} {\bibfnamefont {X.-G.}\ \bibnamefont {Wen}},\ }\href
  {\doibase 10.1103/PhysRevLett.98.117205} {\bibfield  {journal} {\bibinfo
  {journal} {Phys. Rev. Lett.}\ }\textbf {\bibinfo {volume} {98}},\ \bibinfo
  {pages} {117205} (\bibinfo {year} {2007})}\BibitemShut {NoStop}%
\bibitem [{\citenamefont {Wang}\ and\ \citenamefont
  {Vishwanath}(2006)}]{wang06}%
  \BibitemOpen
  \bibfield  {author} {\bibinfo {author} {\bibfnamefont {F.}~\bibnamefont
  {Wang}}\ and\ \bibinfo {author} {\bibfnamefont {A.}~\bibnamefont
  {Vishwanath}},\ }\href {\doibase 10.1103/PhysRevB.74.174423} {\bibfield
  {journal} {\bibinfo  {journal} {Phys. Rev. B}\ }\textbf {\bibinfo {volume}
  {74}},\ \bibinfo {pages} {174423} (\bibinfo {year} {2006})}\BibitemShut
  {NoStop}%
\bibitem [{\citenamefont {Sachdev}(1992)}]{sachdev92}%
  \BibitemOpen
  \bibfield  {author} {\bibinfo {author} {\bibfnamefont {S.}~\bibnamefont
  {Sachdev}},\ }\href {\doibase 10.1103/PhysRevB.45.12377} {\bibfield
  {journal} {\bibinfo  {journal} {Phys. Rev. B}\ }\textbf {\bibinfo {volume}
  {45}},\ \bibinfo {pages} {12377} (\bibinfo {year} {1992})}\BibitemShut
  {NoStop}%
\bibitem [{\citenamefont {Liao}\ \emph {et~al.}(2017)\citenamefont {Liao},
  \citenamefont {Xie}, \citenamefont {Chen}, \citenamefont {Liu}, \citenamefont
  {Xie}, \citenamefont {Huang}, \citenamefont {Normand},\ and\ \citenamefont
  {Xiang}}]{liao17}%
  \BibitemOpen
  \bibfield  {author} {\bibinfo {author} {\bibfnamefont {H.~J.}\ \bibnamefont
  {Liao}}, \bibinfo {author} {\bibfnamefont {Z.~Y.}\ \bibnamefont {Xie}},
  \bibinfo {author} {\bibfnamefont {J.}~\bibnamefont {Chen}}, \bibinfo {author}
  {\bibfnamefont {Z.~Y.}\ \bibnamefont {Liu}}, \bibinfo {author} {\bibfnamefont
  {H.~D.}\ \bibnamefont {Xie}}, \bibinfo {author} {\bibfnamefont {R.~Z.}\
  \bibnamefont {Huang}}, \bibinfo {author} {\bibfnamefont {B.}~\bibnamefont
  {Normand}}, \ and\ \bibinfo {author} {\bibfnamefont {T.}~\bibnamefont
  {Xiang}},\ }\href {\doibase 10.1103/PhysRevLett.118.137202} {\bibfield
  {journal} {\bibinfo  {journal} {Phys. Rev. Lett.}\ }\textbf {\bibinfo
  {volume} {118}},\ \bibinfo {pages} {137202} (\bibinfo {year}
  {2017})}\BibitemShut {NoStop}%
\bibitem [{\citenamefont {Norman}(2016)}]{norman16}%
  \BibitemOpen
  \bibfield  {author} {\bibinfo {author} {\bibfnamefont {M.~R.}\ \bibnamefont
  {Norman}},\ }\href {\doibase 10.1103/RevModPhys.88.041002} {\bibfield
  {journal} {\bibinfo  {journal} {Rev. Mod. Phys.}\ }\textbf {\bibinfo {volume}
  {88}},\ \bibinfo {pages} {041002} (\bibinfo {year} {2016})}\BibitemShut
  {NoStop}%
\bibitem [{\citenamefont {Gong}\ \emph {et~al.}(2013)\citenamefont {Gong},
  \citenamefont {Sheng}, \citenamefont {Motrunich},\ and\ \citenamefont
  {Fisher}}]{gong13}%
  \BibitemOpen
  \bibfield  {author} {\bibinfo {author} {\bibfnamefont {S.-S.}\ \bibnamefont
  {Gong}}, \bibinfo {author} {\bibfnamefont {D.~N.}\ \bibnamefont {Sheng}},
  \bibinfo {author} {\bibfnamefont {O.~I.}\ \bibnamefont {Motrunich}}, \ and\
  \bibinfo {author} {\bibfnamefont {M.~P.~A.}\ \bibnamefont {Fisher}},\ }\href
  {\doibase 10.1103/PhysRevB.88.165138} {\bibfield  {journal} {\bibinfo
  {journal} {Phys. Rev. B}\ }\textbf {\bibinfo {volume} {88}},\ \bibinfo
  {pages} {165138} (\bibinfo {year} {2013})}\BibitemShut {NoStop}%
\bibitem [{\citenamefont {Clark}\ \emph {et~al.}(2011)\citenamefont {Clark},
  \citenamefont {Abanin},\ and\ \citenamefont {Sondhi}}]{clark11}%
  \BibitemOpen
  \bibfield  {author} {\bibinfo {author} {\bibfnamefont {B.~K.}\ \bibnamefont
  {Clark}}, \bibinfo {author} {\bibfnamefont {D.~A.}\ \bibnamefont {Abanin}}, \
  and\ \bibinfo {author} {\bibfnamefont {S.~L.}\ \bibnamefont {Sondhi}},\
  }\href {\doibase 10.1103/PhysRevLett.107.087204} {\bibfield  {journal}
  {\bibinfo  {journal} {Phys. Rev. Lett.}\ }\textbf {\bibinfo {volume} {107}},\
  \bibinfo {pages} {087204} (\bibinfo {year} {2011})}\BibitemShut {NoStop}%
\bibitem [{\citenamefont {Albuquerque}\ \emph {et~al.}(2011)\citenamefont
  {Albuquerque}, \citenamefont {Schwandt}, \citenamefont {Het\'enyi},
  \citenamefont {Capponi}, \citenamefont {Mambrini},\ and\ \citenamefont
  {L\"auchli}}]{albuquerque11}%
  \BibitemOpen
  \bibfield  {author} {\bibinfo {author} {\bibfnamefont {A.~F.}\ \bibnamefont
  {Albuquerque}}, \bibinfo {author} {\bibfnamefont {D.}~\bibnamefont
  {Schwandt}}, \bibinfo {author} {\bibfnamefont {B.}~\bibnamefont {Het\'enyi}},
  \bibinfo {author} {\bibfnamefont {S.}~\bibnamefont {Capponi}}, \bibinfo
  {author} {\bibfnamefont {M.}~\bibnamefont {Mambrini}}, \ and\ \bibinfo
  {author} {\bibfnamefont {A.~M.}\ \bibnamefont {L\"auchli}},\ }\href {\doibase
  10.1103/PhysRevB.84.024406} {\bibfield  {journal} {\bibinfo  {journal} {Phys.
  Rev. B}\ }\textbf {\bibinfo {volume} {84}},\ \bibinfo {pages} {024406}
  (\bibinfo {year} {2011})}\BibitemShut {NoStop}%
\bibitem [{\citenamefont {Ganesh}\ \emph {et~al.}(2013)\citenamefont {Ganesh},
  \citenamefont {van~den Brink},\ and\ \citenamefont {Nishimoto}}]{ganesh13}%
  \BibitemOpen
  \bibfield  {author} {\bibinfo {author} {\bibfnamefont {R.}~\bibnamefont
  {Ganesh}}, \bibinfo {author} {\bibfnamefont {J.}~\bibnamefont {van~den
  Brink}}, \ and\ \bibinfo {author} {\bibfnamefont {S.}~\bibnamefont
  {Nishimoto}},\ }\href@noop {} {\bibfield  {journal} {\bibinfo  {journal}
  {Phys. Rev. Lett.}\ }\textbf {\bibinfo {volume} {110}},\ \bibinfo {pages}
  {127203} (\bibinfo {year} {2013})}\BibitemShut {NoStop}%
\bibitem [{\citenamefont {Zhu}\ \emph {et~al.}(2013)\citenamefont {Zhu},
  \citenamefont {Huse},\ and\ \citenamefont {White}}]{zhu13}%
  \BibitemOpen
  \bibfield  {author} {\bibinfo {author} {\bibfnamefont {Z.}~\bibnamefont
  {Zhu}}, \bibinfo {author} {\bibfnamefont {D.~A.}\ \bibnamefont {Huse}}, \
  and\ \bibinfo {author} {\bibfnamefont {S.~R.}\ \bibnamefont {White}},\ }\href
  {\doibase 10.1103/PhysRevLett.110.127205} {\bibfield  {journal} {\bibinfo
  {journal} {Phys. Rev. Lett.}\ }\textbf {\bibinfo {volume} {110}},\ \bibinfo
  {pages} {127205} (\bibinfo {year} {2013})}\BibitemShut {NoStop}%
\bibitem [{\citenamefont {Ferrari}\ \emph {et~al.}(2017)\citenamefont
  {Ferrari}, \citenamefont {Bieri},\ and\ \citenamefont {Becca}}]{ferrari17}%
  \BibitemOpen
  \bibfield  {author} {\bibinfo {author} {\bibfnamefont {F.}~\bibnamefont
  {Ferrari}}, \bibinfo {author} {\bibfnamefont {S.}~\bibnamefont {Bieri}}, \
  and\ \bibinfo {author} {\bibfnamefont {F.}~\bibnamefont {Becca}},\ }\href
  {\doibase 10.1103/PhysRevB.96.104401} {\bibfield  {journal} {\bibinfo
  {journal} {Phys. Rev. B}\ }\textbf {\bibinfo {volume} {96}},\ \bibinfo
  {pages} {104401} (\bibinfo {year} {2017})}\BibitemShut {NoStop}%
\bibitem [{\citenamefont {Kaneko}\ \emph {et~al.}(2014)\citenamefont {Kaneko},
  \citenamefont {Morita},\ and\ \citenamefont {Imada}}]{kaneko14}%
  \BibitemOpen
  \bibfield  {author} {\bibinfo {author} {\bibfnamefont {R.}~\bibnamefont
  {Kaneko}}, \bibinfo {author} {\bibfnamefont {S.}~\bibnamefont {Morita}}, \
  and\ \bibinfo {author} {\bibfnamefont {M.}~\bibnamefont {Imada}},\ }\href
  {\doibase 10.7566/JPSJ.83.093707} {\bibfield  {journal} {\bibinfo  {journal}
  {Journal of the Physical Society of Japan}\ }\textbf {\bibinfo {volume}
  {83}},\ \bibinfo {pages} {093707} (\bibinfo {year} {2014})},\ \Eprint
  {http://arxiv.org/abs/https://doi.org/10.7566/JPSJ.83.093707}
  {https://doi.org/10.7566/JPSJ.83.093707} \BibitemShut {NoStop}%
\bibitem [{\citenamefont {Shimada}\ \emph
  {et~al.}(2018{\natexlab{a}})\citenamefont {Shimada}, \citenamefont {Nakano},
  \citenamefont {Sakai},\ and\ \citenamefont {Yoshimura}}]{shimada18}%
  \BibitemOpen
  \bibfield  {author} {\bibinfo {author} {\bibfnamefont {A.}~\bibnamefont
  {Shimada}}, \bibinfo {author} {\bibfnamefont {H.}~\bibnamefont {Nakano}},
  \bibinfo {author} {\bibfnamefont {T.}~\bibnamefont {Sakai}}, \ and\ \bibinfo
  {author} {\bibfnamefont {K.}~\bibnamefont {Yoshimura}},\ }\href {\doibase
  10.7566/JPSJ.87.034706} {\bibfield  {journal} {\bibinfo  {journal} {Journal
  of the Physical Society of Japan}\ }\textbf {\bibinfo {volume} {87}},\
  \bibinfo {pages} {034706} (\bibinfo {year} {2018}{\natexlab{a}})},\ \Eprint
  {http://arxiv.org/abs/https://doi.org/10.7566/JPSJ.87.034706}
  {https://doi.org/10.7566/JPSJ.87.034706} \BibitemShut {NoStop}%
\bibitem [{\citenamefont {Saadatmand}\ \emph {et~al.}(2015)\citenamefont
  {Saadatmand}, \citenamefont {Powell},\ and\ \citenamefont
  {McCulloch}}]{saadatmand15}%
  \BibitemOpen
  \bibfield  {author} {\bibinfo {author} {\bibfnamefont {S.~N.}\ \bibnamefont
  {Saadatmand}}, \bibinfo {author} {\bibfnamefont {B.~J.}\ \bibnamefont
  {Powell}}, \ and\ \bibinfo {author} {\bibfnamefont {I.~P.}\ \bibnamefont
  {McCulloch}},\ }\href {\doibase 10.1103/PhysRevB.91.245119} {\bibfield
  {journal} {\bibinfo  {journal} {Phys. Rev. B}\ }\textbf {\bibinfo {volume}
  {91}},\ \bibinfo {pages} {245119} (\bibinfo {year} {2015})}\BibitemShut
  {NoStop}%
\bibitem [{\citenamefont {Itou}\ \emph {et~al.}(2008)\citenamefont {Itou},
  \citenamefont {Oyamada}, \citenamefont {Maegawa}, \citenamefont {Tamura},\
  and\ \citenamefont {Kato}}]{itou08}%
  \BibitemOpen
  \bibfield  {author} {\bibinfo {author} {\bibfnamefont {T.}~\bibnamefont
  {Itou}}, \bibinfo {author} {\bibfnamefont {A.}~\bibnamefont {Oyamada}},
  \bibinfo {author} {\bibfnamefont {S.}~\bibnamefont {Maegawa}}, \bibinfo
  {author} {\bibfnamefont {M.}~\bibnamefont {Tamura}}, \ and\ \bibinfo {author}
  {\bibfnamefont {R.}~\bibnamefont {Kato}},\ }\href@noop {} {\bibfield
  {journal} {\bibinfo  {journal} {Phys. Rev. B}\ }\textbf {\bibinfo {volume}
  {77}},\ \bibinfo {pages} {104413} (\bibinfo {year} {2008})}\BibitemShut
  {NoStop}%
\bibitem [{\citenamefont {Wietek}\ and\ \citenamefont
  {L\"auchli}(2017)}]{wietek17}%
  \BibitemOpen
  \bibfield  {author} {\bibinfo {author} {\bibfnamefont {A.}~\bibnamefont
  {Wietek}}\ and\ \bibinfo {author} {\bibfnamefont {A.~M.}\ \bibnamefont
  {L\"auchli}},\ }\href {\doibase 10.1103/PhysRevB.95.035141} {\bibfield
  {journal} {\bibinfo  {journal} {Phys. Rev. B}\ }\textbf {\bibinfo {volume}
  {95}},\ \bibinfo {pages} {035141} (\bibinfo {year} {2017})}\BibitemShut
  {NoStop}%
\bibitem [{\citenamefont {Iqbal}\ \emph {et~al.}(2016)\citenamefont {Iqbal},
  \citenamefont {Hu}, \citenamefont {Thomale}, \citenamefont {Poilblanc},\ and\
  \citenamefont {Becca}}]{iqbal16}%
  \BibitemOpen
  \bibfield  {author} {\bibinfo {author} {\bibfnamefont {Y.}~\bibnamefont
  {Iqbal}}, \bibinfo {author} {\bibfnamefont {W.-J.}\ \bibnamefont {Hu}},
  \bibinfo {author} {\bibfnamefont {R.}~\bibnamefont {Thomale}}, \bibinfo
  {author} {\bibfnamefont {D.}~\bibnamefont {Poilblanc}}, \ and\ \bibinfo
  {author} {\bibfnamefont {F.}~\bibnamefont {Becca}},\ }\href {\doibase
  10.1103/PhysRevB.93.144411} {\bibfield  {journal} {\bibinfo  {journal} {Phys.
  Rev. B}\ }\textbf {\bibinfo {volume} {93}},\ \bibinfo {pages} {144411}
  (\bibinfo {year} {2016})}\BibitemShut {NoStop}%
\bibitem [{\citenamefont {Hu}\ \emph {et~al.}(2019)\citenamefont {Hu},
  \citenamefont {Zhu}, \citenamefont {Eggert},\ and\ \citenamefont
  {He}}]{shijie19}%
  \BibitemOpen
  \bibfield  {author} {\bibinfo {author} {\bibfnamefont {S.}~\bibnamefont
  {Hu}}, \bibinfo {author} {\bibfnamefont {W.}~\bibnamefont {Zhu}}, \bibinfo
  {author} {\bibfnamefont {S.}~\bibnamefont {Eggert}}, \ and\ \bibinfo {author}
  {\bibfnamefont {Y.-C.}\ \bibnamefont {He}},\ }\href {\doibase
  10.1103/PhysRevLett.123.207203} {\bibfield  {journal} {\bibinfo  {journal}
  {Phys. Rev. Lett.}\ }\textbf {\bibinfo {volume} {123}},\ \bibinfo {pages}
  {207203} (\bibinfo {year} {2019})}\BibitemShut {NoStop}%
\bibitem [{\citenamefont {Sahoo}\ \emph {et~al.}(2018)\citenamefont {Sahoo},
  \citenamefont {Kochkov}, \citenamefont {Clark},\ and\ \citenamefont
  {Flint}}]{js18}%
  \BibitemOpen
  \bibfield  {author} {\bibinfo {author} {\bibfnamefont {J.}~\bibnamefont
  {Sahoo}}, \bibinfo {author} {\bibfnamefont {D.}~\bibnamefont {Kochkov}},
  \bibinfo {author} {\bibfnamefont {B.~K.}\ \bibnamefont {Clark}}, \ and\
  \bibinfo {author} {\bibfnamefont {R.}~\bibnamefont {Flint}},\ }\href
  {\doibase 10.1103/PhysRevB.98.134419} {\bibfield  {journal} {\bibinfo
  {journal} {Phys. Rev. B}\ }\textbf {\bibinfo {volume} {98}},\ \bibinfo
  {pages} {134419} (\bibinfo {year} {2018})}\BibitemShut {NoStop}%
\bibitem [{\citenamefont {Coldea}\ \emph {et~al.}(2001)\citenamefont {Coldea},
  \citenamefont {Tennant}, \citenamefont {Tsvelik},\ and\ \citenamefont
  {Tylczynski}}]{coldea01}%
  \BibitemOpen
  \bibfield  {author} {\bibinfo {author} {\bibfnamefont {R.}~\bibnamefont
  {Coldea}}, \bibinfo {author} {\bibfnamefont {D.~A.}\ \bibnamefont {Tennant}},
  \bibinfo {author} {\bibfnamefont {A.~M.}\ \bibnamefont {Tsvelik}}, \ and\
  \bibinfo {author} {\bibfnamefont {Z.}~\bibnamefont {Tylczynski}},\ }\href
  {\doibase 10.1103/PhysRevLett.86.1335} {\bibfield  {journal} {\bibinfo
  {journal} {Phys. Rev. Lett.}\ }\textbf {\bibinfo {volume} {86}},\ \bibinfo
  {pages} {1335} (\bibinfo {year} {2001})}\BibitemShut {NoStop}%
\bibitem [{\citenamefont {Shimizu}\ \emph {et~al.}(2003)\citenamefont
  {Shimizu}, \citenamefont {Miyagawa}, \citenamefont {Kanoda}, \citenamefont
  {Maesato},\ and\ \citenamefont {Saito}}]{shimizu03}%
  \BibitemOpen
  \bibfield  {author} {\bibinfo {author} {\bibfnamefont {Y.}~\bibnamefont
  {Shimizu}}, \bibinfo {author} {\bibfnamefont {K.}~\bibnamefont {Miyagawa}},
  \bibinfo {author} {\bibfnamefont {K.}~\bibnamefont {Kanoda}}, \bibinfo
  {author} {\bibfnamefont {M.}~\bibnamefont {Maesato}}, \ and\ \bibinfo
  {author} {\bibfnamefont {G.}~\bibnamefont {Saito}},\ }\href@noop {}
  {\bibfield  {journal} {\bibinfo  {journal} {Phys. Rev. Lett.}\ }\textbf
  {\bibinfo {volume} {91}},\ \bibinfo {pages} {107001} (\bibinfo {year}
  {2003})}\BibitemShut {NoStop}%
\bibitem [{\citenamefont {Yamashita}\ \emph {et~al.}(2009)\citenamefont
  {Yamashita}, \citenamefont {Nakata}, \citenamefont {Kasahara}, \citenamefont
  {Sasaki}, \citenamefont {Yoneyama}, \citenamefont {Kobayashi}, \citenamefont
  {Fujimoto}, \citenamefont {Shibauchi},\ and\ \citenamefont
  {Matsuda}}]{yamashita08}%
  \BibitemOpen
  \bibfield  {author} {\bibinfo {author} {\bibfnamefont {M.}~\bibnamefont
  {Yamashita}}, \bibinfo {author} {\bibfnamefont {N.}~\bibnamefont {Nakata}},
  \bibinfo {author} {\bibfnamefont {Y.}~\bibnamefont {Kasahara}}, \bibinfo
  {author} {\bibfnamefont {T.}~\bibnamefont {Sasaki}}, \bibinfo {author}
  {\bibfnamefont {N.}~\bibnamefont {Yoneyama}}, \bibinfo {author}
  {\bibfnamefont {N.}~\bibnamefont {Kobayashi}}, \bibinfo {author}
  {\bibfnamefont {S.}~\bibnamefont {Fujimoto}}, \bibinfo {author}
  {\bibfnamefont {T.}~\bibnamefont {Shibauchi}}, \ and\ \bibinfo {author}
  {\bibfnamefont {Y.}~\bibnamefont {Matsuda}},\ }\href@noop {} {\bibfield
  {journal} {\bibinfo  {journal} {Nature Physics}\ }\textbf {\bibinfo {volume}
  {5}},\ \bibinfo {pages} {44} (\bibinfo {year} {2009})}\BibitemShut {NoStop}%
\bibitem [{\citenamefont {Chubukov}\ and\ \citenamefont
  {Starykh}(2013)}]{chubukov13}%
  \BibitemOpen
  \bibfield  {author} {\bibinfo {author} {\bibfnamefont {A.~V.}\ \bibnamefont
  {Chubukov}}\ and\ \bibinfo {author} {\bibfnamefont {O.~A.}\ \bibnamefont
  {Starykh}},\ }\href {\doibase 10.1103/PhysRevLett.110.217210} {\bibfield
  {journal} {\bibinfo  {journal} {Phys. Rev. Lett.}\ }\textbf {\bibinfo
  {volume} {110}},\ \bibinfo {pages} {217210} (\bibinfo {year}
  {2013})}\BibitemShut {NoStop}%
\bibitem [{\citenamefont {Starykh}(2015)}]{starykh15}%
  \BibitemOpen
  \bibfield  {author} {\bibinfo {author} {\bibfnamefont {O.~A.}\ \bibnamefont
  {Starykh}},\ }\href {\doibase https://doi.org/10.1088/0034-4885/78/5/052502}
  {\bibfield  {journal} {\bibinfo  {journal} {Rep. Prog. Phys.}\ }\textbf
  {\bibinfo {volume} {78}},\ \bibinfo {pages} {052502} (\bibinfo {year}
  {2015})}\BibitemShut {NoStop}%
\bibitem [{\citenamefont {McKenzie}(1998)}]{mckenzie98}%
  \BibitemOpen
  \bibfield  {author} {\bibinfo {author} {\bibfnamefont {R.~H.}\ \bibnamefont
  {McKenzie}},\ }\href@noop {} {\  (\bibinfo {year} {1998})},\ \Eprint
  {http://arxiv.org/abs/cond-mat/9802198} {arXiv:cond-mat/9802198
  [cond-mat.str-el]} \BibitemShut {NoStop}%
\bibitem [{\citenamefont {Nakano}\ and\ \citenamefont
  {Sakai}(2017)}]{nakano17}%
  \BibitemOpen
  \bibfield  {author} {\bibinfo {author} {\bibfnamefont {H.}~\bibnamefont
  {Nakano}}\ and\ \bibinfo {author} {\bibfnamefont {T.}~\bibnamefont {Sakai}},\
  }\href {\doibase 10.7566/JPSJ.86.063702} {\bibfield  {journal} {\bibinfo
  {journal} {Journal of the Physical Society of Japan}\ }\textbf {\bibinfo
  {volume} {86}},\ \bibinfo {pages} {063702} (\bibinfo {year} {2017})},\
  \Eprint {http://arxiv.org/abs/https://doi.org/10.7566/JPSJ.86.063702}
  {https://doi.org/10.7566/JPSJ.86.063702} \BibitemShut {NoStop}%
\bibitem [{\citenamefont {Gonzalez}\ \emph {et~al.}(2018)\citenamefont
  {Gonzalez}, \citenamefont {Lisandrini}, \citenamefont {Blesio}, \citenamefont
  {Trumper}, \citenamefont {Gazza},\ and\ \citenamefont {Manuel}}]{gonzalez18}%
  \BibitemOpen
  \bibfield  {author} {\bibinfo {author} {\bibfnamefont {M.}~\bibnamefont
  {Gonzalez}}, \bibinfo {author} {\bibfnamefont {F.~T.}\ \bibnamefont
  {Lisandrini}}, \bibinfo {author} {\bibfnamefont {G.~G.}\ \bibnamefont
  {Blesio}}, \bibinfo {author} {\bibfnamefont {A.~E.}\ \bibnamefont {Trumper}},
  \bibinfo {author} {\bibfnamefont {C.~J.}\ \bibnamefont {Gazza}}, \ and\
  \bibinfo {author} {\bibfnamefont {L.~O.}\ \bibnamefont {Manuel}},\
  }\href@noop {} {\bibfield  {journal} {\bibinfo  {journal} {ArXiv e-prints}\ }
  (\bibinfo {year} {2018})},\ \Eprint {http://arxiv.org/abs/1804.06720}
  {arXiv:1804.06720 [cond-mat.str-el]} \BibitemShut {NoStop}%
\bibitem [{\citenamefont {Shimada}\ \emph
  {et~al.}(2018{\natexlab{b}})\citenamefont {Shimada}, \citenamefont {Sakai},
  \citenamefont {Nakano},\ and\ \citenamefont {Yoshimura}}]{shimada182}%
  \BibitemOpen
  \bibfield  {author} {\bibinfo {author} {\bibfnamefont {A.}~\bibnamefont
  {Shimada}}, \bibinfo {author} {\bibfnamefont {T.}~\bibnamefont {Sakai}},
  \bibinfo {author} {\bibfnamefont {H.}~\bibnamefont {Nakano}}, \ and\ \bibinfo
  {author} {\bibfnamefont {K.}~\bibnamefont {Yoshimura}},\ }\href
  {http://stacks.iop.org/1742-6596/969/i=1/a=012126} {\bibfield  {journal}
  {\bibinfo  {journal} {Journal of Physics: Conference Series}\ }\textbf
  {\bibinfo {volume} {969}},\ \bibinfo {pages} {012126} (\bibinfo {year}
  {2018}{\natexlab{b}})}\BibitemShut {NoStop}%
\bibitem [{\citenamefont {Seifert}\ and\ \citenamefont
  {Vojta}(2019)}]{siefert19}%
  \BibitemOpen
  \bibfield  {author} {\bibinfo {author} {\bibfnamefont {U.~F.~P.}\
  \bibnamefont {Seifert}}\ and\ \bibinfo {author} {\bibfnamefont
  {M.}~\bibnamefont {Vojta}},\ }\href {\doibase 10.1103/PhysRevB.99.155156}
  {\bibfield  {journal} {\bibinfo  {journal} {Phys. Rev. B}\ }\textbf {\bibinfo
  {volume} {99}},\ \bibinfo {pages} {155156} (\bibinfo {year}
  {2019})}\BibitemShut {NoStop}%
\bibitem [{\citenamefont {Gordon}\ \emph {et~al.}(2018)\citenamefont {Gordon},
  \citenamefont {Cheng}, \citenamefont {Kim}, \citenamefont {Cheong},
  \citenamefont {Deng},\ and\ \citenamefont {Whangbo}}]{gordon18}%
  \BibitemOpen
  \bibfield  {author} {\bibinfo {author} {\bibfnamefont {E.~E.}\ \bibnamefont
  {Gordon}}, \bibinfo {author} {\bibfnamefont {X.}~\bibnamefont {Cheng}},
  \bibinfo {author} {\bibfnamefont {J.}~\bibnamefont {Kim}}, \bibinfo {author}
  {\bibfnamefont {S.-W.}\ \bibnamefont {Cheong}}, \bibinfo {author}
  {\bibfnamefont {S.}~\bibnamefont {Deng}}, \ and\ \bibinfo {author}
  {\bibfnamefont {M.-H.}\ \bibnamefont {Whangbo}},\ }\href {\doibase
  10.1021/acs.inorgchem.8b01274} {\bibfield  {journal} {\bibinfo  {journal}
  {Inorganic Chemistry}\ }\textbf {\bibinfo {volume} {57}},\ \bibinfo {pages}
  {9260} (\bibinfo {year} {2018})},\ \bibinfo {note} {pMID: 30036040},\ \Eprint
  {http://arxiv.org/abs/https://doi.org/10.1021/acs.inorgchem.8b01274}
  {https://doi.org/10.1021/acs.inorgchem.8b01274} \BibitemShut {NoStop}%
\bibitem [{\citenamefont {Chen}\ \emph {et~al.}(2018)\citenamefont {Chen},
  \citenamefont {Holinsworth}, \citenamefont {O’Neal}, \citenamefont {Luo},
  \citenamefont {Topping}, \citenamefont {Cheong}, \citenamefont {Singleton},
  \citenamefont {Choi},\ and\ \citenamefont {Musfeldt}}]{peng18}%
  \BibitemOpen
  \bibfield  {author} {\bibinfo {author} {\bibfnamefont {P.}~\bibnamefont
  {Chen}}, \bibinfo {author} {\bibfnamefont {B.~S.}\ \bibnamefont
  {Holinsworth}}, \bibinfo {author} {\bibfnamefont {K.~R.}\ \bibnamefont
  {O’Neal}}, \bibinfo {author} {\bibfnamefont {X.}~\bibnamefont {Luo}},
  \bibinfo {author} {\bibfnamefont {C.~V.}\ \bibnamefont {Topping}}, \bibinfo
  {author} {\bibfnamefont {S.~W.}\ \bibnamefont {Cheong}}, \bibinfo {author}
  {\bibfnamefont {J.}~\bibnamefont {Singleton}}, \bibinfo {author}
  {\bibfnamefont {E.~S.}\ \bibnamefont {Choi}}, \ and\ \bibinfo {author}
  {\bibfnamefont {J.~L.}\ \bibnamefont {Musfeldt}},\ }\href {\doibase
  10.1021/acs.inorgchem.8b01467} {\bibfield  {journal} {\bibinfo  {journal}
  {Inorganic Chemistry}\ }\textbf {\bibinfo {volume} {57}},\ \bibinfo {pages}
  {12501} (\bibinfo {year} {2018})},\ \Eprint
  {http://arxiv.org/abs/https://doi.org/10.1021/acs.inorgchem.8b01467}
  {https://doi.org/10.1021/acs.inorgchem.8b01467} \BibitemShut {NoStop}%
\bibitem [{\citenamefont {Clark}\ \emph {et~al.}(2019)\citenamefont {Clark},
  \citenamefont {Sala}, \citenamefont {Maharaj}, \citenamefont {Stone},
  \citenamefont {Knight}, \citenamefont {Telling}, \citenamefont {Wang},
  \citenamefont {Xu}, \citenamefont {Kim}, \citenamefont {Yanbin},
  \citenamefont {Cheong},\ and\ \citenamefont {Gaulin}}]{clark19}%
  \BibitemOpen
  \bibfield  {author} {\bibinfo {author} {\bibfnamefont {L.}~\bibnamefont
  {Clark}}, \bibinfo {author} {\bibfnamefont {G.}~\bibnamefont {Sala}},
  \bibinfo {author} {\bibfnamefont {D.~D.}\ \bibnamefont {Maharaj}}, \bibinfo
  {author} {\bibfnamefont {M.~B.}\ \bibnamefont {Stone}}, \bibinfo {author}
  {\bibfnamefont {K.~S.}\ \bibnamefont {Knight}}, \bibinfo {author}
  {\bibfnamefont {M.~T.~F.}\ \bibnamefont {Telling}}, \bibinfo {author}
  {\bibfnamefont {X.}~\bibnamefont {Wang}}, \bibinfo {author} {\bibfnamefont
  {X.}~\bibnamefont {Xu}}, \bibinfo {author} {\bibfnamefont {J.}~\bibnamefont
  {Kim}}, \bibinfo {author} {\bibfnamefont {L.}~\bibnamefont {Yanbin}},
  \bibinfo {author} {\bibfnamefont {S.-W.}\ \bibnamefont {Cheong}}, \ and\
  \bibinfo {author} {\bibfnamefont {B.~D.}\ \bibnamefont {Gaulin}},\ }\href
  {\doibase https://doi.org/10.1038/s41567-018-0407-2} {\bibfield  {journal}
  {\bibinfo  {journal} {Nature Physics}\ }\textbf {\bibinfo {volume} {15}},\
  \bibinfo {pages} {262} (\bibinfo {year} {2019})}\BibitemShut {NoStop}%
\bibitem [{\citenamefont {Kim}\ \emph {et~al.}(2019{\natexlab{a}})\citenamefont
  {Kim}, \citenamefont {Wang}, \citenamefont {Huang}, \citenamefont {Wang},
  \citenamefont {Fang}, \citenamefont {Luo}, \citenamefont {Li}, \citenamefont
  {Wu}, \citenamefont {Mori}, \citenamefont {Kwok}, \citenamefont {Mun},
  \citenamefont {Zapf},\ and\ \citenamefont {Cheong}}]{kim19}%
  \BibitemOpen
  \bibfield  {author} {\bibinfo {author} {\bibfnamefont {J.}~\bibnamefont
  {Kim}}, \bibinfo {author} {\bibfnamefont {X.}~\bibnamefont {Wang}}, \bibinfo
  {author} {\bibfnamefont {F.-T.}\ \bibnamefont {Huang}}, \bibinfo {author}
  {\bibfnamefont {Y.}~\bibnamefont {Wang}}, \bibinfo {author} {\bibfnamefont
  {X.}~\bibnamefont {Fang}}, \bibinfo {author} {\bibfnamefont {X.}~\bibnamefont
  {Luo}}, \bibinfo {author} {\bibfnamefont {Y.}~\bibnamefont {Li}}, \bibinfo
  {author} {\bibfnamefont {M.}~\bibnamefont {Wu}}, \bibinfo {author}
  {\bibfnamefont {S.}~\bibnamefont {Mori}}, \bibinfo {author} {\bibfnamefont
  {D.}~\bibnamefont {Kwok}}, \bibinfo {author} {\bibfnamefont {E.~D.}\
  \bibnamefont {Mun}}, \bibinfo {author} {\bibfnamefont {V.~S.}\ \bibnamefont
  {Zapf}}, \ and\ \bibinfo {author} {\bibfnamefont {S.-W.}\ \bibnamefont
  {Cheong}},\ }\href {\doibase 10.1103/PhysRevX.9.031005} {\bibfield  {journal}
  {\bibinfo  {journal} {Phys. Rev. X}\ }\textbf {\bibinfo {volume} {9}},\
  \bibinfo {pages} {031005} (\bibinfo {year} {2019}{\natexlab{a}})}\BibitemShut
  {NoStop}%
\bibitem [{\citenamefont {Kim}\ \emph {et~al.}(2019{\natexlab{b}})\citenamefont
  {Kim}, \citenamefont {Winn}, \citenamefont {Chi}, \citenamefont {Savici},
  \citenamefont {Rodriguez-Rivera}, \citenamefont {Chen}, \citenamefont {Xu},
  \citenamefont {Li}, \citenamefont {Kim}, \citenamefont {Cheong},\ and\
  \citenamefont {Kiryukhin}}]{kimprb19}%
  \BibitemOpen
  \bibfield  {author} {\bibinfo {author} {\bibfnamefont {M.~G.}\ \bibnamefont
  {Kim}}, \bibinfo {author} {\bibfnamefont {B.}~\bibnamefont {Winn}}, \bibinfo
  {author} {\bibfnamefont {S.}~\bibnamefont {Chi}}, \bibinfo {author}
  {\bibfnamefont {A.~T.}\ \bibnamefont {Savici}}, \bibinfo {author}
  {\bibfnamefont {J.~A.}\ \bibnamefont {Rodriguez-Rivera}}, \bibinfo {author}
  {\bibfnamefont {W.~C.}\ \bibnamefont {Chen}}, \bibinfo {author}
  {\bibfnamefont {X.}~\bibnamefont {Xu}}, \bibinfo {author} {\bibfnamefont
  {Y.}~\bibnamefont {Li}}, \bibinfo {author} {\bibfnamefont {J.~W.}\
  \bibnamefont {Kim}}, \bibinfo {author} {\bibfnamefont {S.-W.}\ \bibnamefont
  {Cheong}}, \ and\ \bibinfo {author} {\bibfnamefont {V.}~\bibnamefont
  {Kiryukhin}},\ }\href {\doibase 10.1103/PhysRevB.100.024405} {\bibfield
  {journal} {\bibinfo  {journal} {Phys. Rev. B}\ }\textbf {\bibinfo {volume}
  {100}},\ \bibinfo {pages} {024405} (\bibinfo {year}
  {2019}{\natexlab{b}})}\BibitemShut {NoStop}%
\bibitem [{\citenamefont {Flint}\ and\ \citenamefont {Lee}(2013)}]{flint13}%
  \BibitemOpen
  \bibfield  {author} {\bibinfo {author} {\bibfnamefont {R.}~\bibnamefont
  {Flint}}\ and\ \bibinfo {author} {\bibfnamefont {P.~A.}\ \bibnamefont
  {Lee}},\ }\href {\doibase 10.1103/PhysRevLett.111.217201} {\bibfield
  {journal} {\bibinfo  {journal} {Phys. Rev. Lett.}\ }\textbf {\bibinfo
  {volume} {111}},\ \bibinfo {pages} {217201} (\bibinfo {year}
  {2013})}\BibitemShut {NoStop}%
\bibitem [{\citenamefont {Sheckelton}\ \emph {et~al.}(2012)\citenamefont
  {Sheckelton}, \citenamefont {Neilson}, \citenamefont {Soltan},\ and\
  \citenamefont {McQueen}}]{sheckelton12}%
  \BibitemOpen
  \bibfield  {author} {\bibinfo {author} {\bibfnamefont {J.~P.}\ \bibnamefont
  {Sheckelton}}, \bibinfo {author} {\bibfnamefont {J.~R.}\ \bibnamefont
  {Neilson}}, \bibinfo {author} {\bibfnamefont {D.~G.}\ \bibnamefont {Soltan}},
  \ and\ \bibinfo {author} {\bibfnamefont {T.~M.}\ \bibnamefont {McQueen}},\
  }\href@noop {} {\bibfield  {journal} {\bibinfo  {journal} {Nature Materials}\
  }\textbf {\bibinfo {volume} {11}},\ \bibinfo {pages} {493} (\bibinfo {year}
  {2012})}\BibitemShut {NoStop}%
\bibitem [{\citenamefont {Mourigal}\ \emph {et~al.}(2014)\citenamefont
  {Mourigal}, \citenamefont {Fuhrman}, \citenamefont {Sheckelton},
  \citenamefont {Wartelle}, \citenamefont {Rodriguez-Rivera}, \citenamefont
  {Abernathy}, \citenamefont {McQueen},\ and\ \citenamefont
  {Broholm}}]{mourigal14}%
  \BibitemOpen
  \bibfield  {author} {\bibinfo {author} {\bibfnamefont {M.}~\bibnamefont
  {Mourigal}}, \bibinfo {author} {\bibfnamefont {W.~T.}\ \bibnamefont
  {Fuhrman}}, \bibinfo {author} {\bibfnamefont {J.~P.}\ \bibnamefont
  {Sheckelton}}, \bibinfo {author} {\bibfnamefont {A.}~\bibnamefont
  {Wartelle}}, \bibinfo {author} {\bibfnamefont {J.~A.}\ \bibnamefont
  {Rodriguez-Rivera}}, \bibinfo {author} {\bibfnamefont {D.~L.}\ \bibnamefont
  {Abernathy}}, \bibinfo {author} {\bibfnamefont {T.~M.}\ \bibnamefont
  {McQueen}}, \ and\ \bibinfo {author} {\bibfnamefont {C.~L.}\ \bibnamefont
  {Broholm}},\ }\href {\doibase 10.1103/PhysRevLett.112.027202} {\bibfield
  {journal} {\bibinfo  {journal} {Phys. Rev. Lett.}\ }\textbf {\bibinfo
  {volume} {112}},\ \bibinfo {pages} {027202} (\bibinfo {year}
  {2014})}\BibitemShut {NoStop}%
\bibitem [{\citenamefont {Sheckelton}\ \emph {et~al.}(2014)\citenamefont
  {Sheckelton}, \citenamefont {Foronda}, \citenamefont {Pan}, \citenamefont
  {Moir}, \citenamefont {McDonald}, \citenamefont {Lancaster}, \citenamefont
  {Baker}, \citenamefont {Armitage}, \citenamefont {Imai}, \citenamefont
  {Blundell},\ and\ \citenamefont {McQueen}}]{sheckelton14}%
  \BibitemOpen
  \bibfield  {author} {\bibinfo {author} {\bibfnamefont {J.~P.}\ \bibnamefont
  {Sheckelton}}, \bibinfo {author} {\bibfnamefont {F.~R.}\ \bibnamefont
  {Foronda}}, \bibinfo {author} {\bibfnamefont {L.~D.}\ \bibnamefont {Pan}},
  \bibinfo {author} {\bibfnamefont {C.}~\bibnamefont {Moir}}, \bibinfo {author}
  {\bibfnamefont {R.~D.}\ \bibnamefont {McDonald}}, \bibinfo {author}
  {\bibfnamefont {T.}~\bibnamefont {Lancaster}}, \bibinfo {author}
  {\bibfnamefont {P.~J.}\ \bibnamefont {Baker}}, \bibinfo {author}
  {\bibfnamefont {N.~P.}\ \bibnamefont {Armitage}}, \bibinfo {author}
  {\bibfnamefont {T.}~\bibnamefont {Imai}}, \bibinfo {author} {\bibfnamefont
  {S.~J.}\ \bibnamefont {Blundell}}, \ and\ \bibinfo {author} {\bibfnamefont
  {T.~M.}\ \bibnamefont {McQueen}},\ }\href {\doibase
  10.1103/PhysRevB.89.064407} {\bibfield  {journal} {\bibinfo  {journal} {Phys.
  Rev. B}\ }\textbf {\bibinfo {volume} {89}},\ \bibinfo {pages} {064407}
  (\bibinfo {year} {2014})}\BibitemShut {NoStop}%
\bibitem [{\citenamefont {Sheckelton}\ \emph {et~al.}(2015)\citenamefont
  {Sheckelton}, \citenamefont {Neilsonab},\ and\ \citenamefont
  {McQueen}}]{sheckelton15}%
  \BibitemOpen
  \bibfield  {author} {\bibinfo {author} {\bibfnamefont {J.~P.}\ \bibnamefont
  {Sheckelton}}, \bibinfo {author} {\bibfnamefont {J.~R.}\ \bibnamefont
  {Neilsonab}}, \ and\ \bibinfo {author} {\bibfnamefont {T.~M.}\ \bibnamefont
  {McQueen}},\ }\href@noop {} {\bibfield  {journal} {\bibinfo  {journal}
  {Mater. Horiz.}\ }\textbf {\bibinfo {volume} {2}},\ \bibinfo {pages} {76}
  (\bibinfo {year} {2015})}\BibitemShut {NoStop}%
\bibitem [{\citenamefont {Chen}\ \emph {et~al.}(2016)\citenamefont {Chen},
  \citenamefont {Kee},\ and\ \citenamefont {Kim}}]{chen16}%
  \BibitemOpen
  \bibfield  {author} {\bibinfo {author} {\bibfnamefont {G.}~\bibnamefont
  {Chen}}, \bibinfo {author} {\bibfnamefont {H.-Y.}\ \bibnamefont {Kee}}, \
  and\ \bibinfo {author} {\bibfnamefont {Y.~B.}\ \bibnamefont {Kim}},\ }\href
  {\doibase 10.1103/PhysRevB.93.245134} {\bibfield  {journal} {\bibinfo
  {journal} {Phys. Rev. B}\ }\textbf {\bibinfo {volume} {93}},\ \bibinfo
  {pages} {245134} (\bibinfo {year} {2016})}\BibitemShut {NoStop}%
\bibitem [{\citenamefont {Carrasquilla}\ \emph {et~al.}(2017)\citenamefont
  {Carrasquilla}, \citenamefont {Chen},\ and\ \citenamefont
  {Melko}}]{carrasquilla17}%
  \BibitemOpen
  \bibfield  {author} {\bibinfo {author} {\bibfnamefont {J.}~\bibnamefont
  {Carrasquilla}}, \bibinfo {author} {\bibfnamefont {G.}~\bibnamefont {Chen}},
  \ and\ \bibinfo {author} {\bibfnamefont {R.~G.}\ \bibnamefont {Melko}},\
  }\href@noop {} {\bibfield  {journal} {\bibinfo  {journal} {Phys. Rev. B}\
  }\textbf {\bibinfo {volume} {96}},\ \bibinfo {pages} {054405} (\bibinfo
  {year} {2017})}\BibitemShut {NoStop}%
\bibitem [{\citenamefont {Chen}\ and\ \citenamefont {Lee}(2018)}]{chen18}%
  \BibitemOpen
  \bibfield  {author} {\bibinfo {author} {\bibfnamefont {G.}~\bibnamefont
  {Chen}}\ and\ \bibinfo {author} {\bibfnamefont {P.~A.}\ \bibnamefont {Lee}},\
  }\href {\doibase 10.1103/PhysRevB.97.035124} {\bibfield  {journal} {\bibinfo
  {journal} {Phys. Rev. B}\ }\textbf {\bibinfo {volume} {97}},\ \bibinfo
  {pages} {035124} (\bibinfo {year} {2018})}\BibitemShut {NoStop}%
\bibitem [{\citenamefont {Wen}(2002)}]{wen02}%
  \BibitemOpen
  \bibfield  {author} {\bibinfo {author} {\bibfnamefont {X.}~\bibnamefont
  {Wen}},\ }\href@noop {} {\bibfield  {journal} {\bibinfo  {journal} {Phys.
  Rev. B.}\ }\textbf {\bibinfo {volume} {65}},\ \bibinfo {pages} {165113}
  (\bibinfo {year} {2002})}\BibitemShut {NoStop}%
\bibitem [{\citenamefont {Ghorbani}\ \emph {et~al.}(2016)\citenamefont
  {Ghorbani}, \citenamefont {Tocchio},\ and\ \citenamefont
  {Becca}}]{ghorbani16}%
  \BibitemOpen
  \bibfield  {author} {\bibinfo {author} {\bibfnamefont {E.}~\bibnamefont
  {Ghorbani}}, \bibinfo {author} {\bibfnamefont {L.~F.}\ \bibnamefont
  {Tocchio}}, \ and\ \bibinfo {author} {\bibfnamefont {F.}~\bibnamefont
  {Becca}},\ }\href {\doibase 10.1103/PhysRevB.93.085111} {\bibfield  {journal}
  {\bibinfo  {journal} {Phys. Rev. B}\ }\textbf {\bibinfo {volume} {93}},\
  \bibinfo {pages} {085111} (\bibinfo {year} {2016})}\BibitemShut {NoStop}%
\bibitem [{\citenamefont {Bieri}\ \emph {et~al.}(2016)\citenamefont {Bieri},
  \citenamefont {Lhuillier},\ and\ \citenamefont {Messio}}]{bieri16}%
  \BibitemOpen
  \bibfield  {author} {\bibinfo {author} {\bibfnamefont {S.}~\bibnamefont
  {Bieri}}, \bibinfo {author} {\bibfnamefont {C.}~\bibnamefont {Lhuillier}}, \
  and\ \bibinfo {author} {\bibfnamefont {L.}~\bibnamefont {Messio}},\ }\href
  {\doibase 10.1103/PhysRevB.93.094437} {\bibfield  {journal} {\bibinfo
  {journal} {Phys. Rev. B}\ }\textbf {\bibinfo {volume} {93}},\ \bibinfo
  {pages} {094437} (\bibinfo {year} {2016})}\BibitemShut {NoStop}%
\bibitem [{\citenamefont {Lu}(2016)}]{lu16}%
  \BibitemOpen
  \bibfield  {author} {\bibinfo {author} {\bibfnamefont {Y.-M.}\ \bibnamefont
  {Lu}},\ }\href {\doibase 10.1103/PhysRevB.93.165113} {\bibfield  {journal}
  {\bibinfo  {journal} {Phys. Rev. B}\ }\textbf {\bibinfo {volume} {93}},\
  \bibinfo {pages} {165113} (\bibinfo {year} {2016})}\BibitemShut {NoStop}%
\bibitem [{\citenamefont {Lu}\ and\ \citenamefont {Ran}(2011)}]{lu11}%
  \BibitemOpen
  \bibfield  {author} {\bibinfo {author} {\bibfnamefont {Y.-M.}\ \bibnamefont
  {Lu}}\ and\ \bibinfo {author} {\bibfnamefont {Y.}~\bibnamefont {Ran}},\
  }\href {\doibase 10.1103/PhysRevB.84.024420} {\bibfield  {journal} {\bibinfo
  {journal} {Phys. Rev. B}\ }\textbf {\bibinfo {volume} {84}},\ \bibinfo
  {pages} {024420} (\bibinfo {year} {2011})}\BibitemShut {NoStop}%
\bibitem [{\citenamefont {Flint}\ and\ \citenamefont
  {Coleman}(2012)}]{flint12}%
  \BibitemOpen
  \bibfield  {author} {\bibinfo {author} {\bibfnamefont {R.}~\bibnamefont
  {Flint}}\ and\ \bibinfo {author} {\bibfnamefont {P.}~\bibnamefont
  {Coleman}},\ }\href {\doibase 10.1103/PhysRevB.86.184508} {\bibfield
  {journal} {\bibinfo  {journal} {Phys. Rev. B}\ }\textbf {\bibinfo {volume}
  {86}},\ \bibinfo {pages} {184508} (\bibinfo {year} {2012})}\BibitemShut
  {NoStop}%
\bibitem [{\citenamefont {Hermele}\ \emph {et~al.}(2004)\citenamefont
  {Hermele}, \citenamefont {Senthil}, \citenamefont {Fisher}, \citenamefont
  {Lee}, \citenamefont {Nagaosa},\ and\ \citenamefont {Wen}}]{hermele04}%
  \BibitemOpen
  \bibfield  {author} {\bibinfo {author} {\bibfnamefont {M.}~\bibnamefont
  {Hermele}}, \bibinfo {author} {\bibfnamefont {T.}~\bibnamefont {Senthil}},
  \bibinfo {author} {\bibfnamefont {M.~P.~A.}\ \bibnamefont {Fisher}}, \bibinfo
  {author} {\bibfnamefont {P.~A.}\ \bibnamefont {Lee}}, \bibinfo {author}
  {\bibfnamefont {N.}~\bibnamefont {Nagaosa}}, \ and\ \bibinfo {author}
  {\bibfnamefont {X.-G.}\ \bibnamefont {Wen}},\ }\href {\doibase
  10.1103/PhysRevB.70.214437} {\bibfield  {journal} {\bibinfo  {journal} {Phys.
  Rev. B}\ }\textbf {\bibinfo {volume} {70}},\ \bibinfo {pages} {214437}
  (\bibinfo {year} {2004})}\BibitemShut {NoStop}%
\bibitem [{\citenamefont {Lee}\ \emph {et~al.}(2006)\citenamefont {Lee},
  \citenamefont {Nagaosa},\ and\ \citenamefont {Wen}}]{lee06}%
  \BibitemOpen
  \bibfield  {author} {\bibinfo {author} {\bibfnamefont {P.~A.}\ \bibnamefont
  {Lee}}, \bibinfo {author} {\bibfnamefont {N.}~\bibnamefont {Nagaosa}}, \ and\
  \bibinfo {author} {\bibfnamefont {X.-G.}\ \bibnamefont {Wen}},\ }\href
  {\doibase 10.1103/RevModPhys.78.17} {\bibfield  {journal} {\bibinfo
  {journal} {Rev. Mod. Phys.}\ }\textbf {\bibinfo {volume} {78}},\ \bibinfo
  {pages} {17} (\bibinfo {year} {2006})}\BibitemShut {NoStop}%
\bibitem [{\citenamefont {Motrunich}(2005)}]{motrunich05}%
  \BibitemOpen
  \bibfield  {author} {\bibinfo {author} {\bibfnamefont {O.~I.}\ \bibnamefont
  {Motrunich}},\ }\href {\doibase 10.1103/PhysRevB.72.045105} {\bibfield
  {journal} {\bibinfo  {journal} {Phys. Rev. B}\ }\textbf {\bibinfo {volume}
  {72}},\ \bibinfo {pages} {045105} (\bibinfo {year} {2005})}\BibitemShut
  {NoStop}%
\bibitem [{\citenamefont {He}\ \emph {et~al.}(2018)\citenamefont {He},
  \citenamefont {Xu}, \citenamefont {Chen}, \citenamefont {Law},\ and\
  \citenamefont {Lee}}]{wen18}%
  \BibitemOpen
  \bibfield  {author} {\bibinfo {author} {\bibfnamefont {W.-Y.}\ \bibnamefont
  {He}}, \bibinfo {author} {\bibfnamefont {X.~Y.}\ \bibnamefont {Xu}}, \bibinfo
  {author} {\bibfnamefont {G.}~\bibnamefont {Chen}}, \bibinfo {author}
  {\bibfnamefont {K.~T.}\ \bibnamefont {Law}}, \ and\ \bibinfo {author}
  {\bibfnamefont {P.~A.}\ \bibnamefont {Lee}},\ }\href {\doibase
  10.1103/PhysRevLett.121.046401} {\bibfield  {journal} {\bibinfo  {journal}
  {Phys. Rev. Lett.}\ }\textbf {\bibinfo {volume} {121}},\ \bibinfo {pages}
  {046401} (\bibinfo {year} {2018})}\BibitemShut {NoStop}%
\bibitem [{\citenamefont {Mishmash}\ \emph {et~al.}(2013)\citenamefont
  {Mishmash}, \citenamefont {Garrison}, \citenamefont {Bieri},\ and\
  \citenamefont {Xu}}]{mishmash13}%
  \BibitemOpen
  \bibfield  {author} {\bibinfo {author} {\bibfnamefont {R.~V.}\ \bibnamefont
  {Mishmash}}, \bibinfo {author} {\bibfnamefont {J.~R.}\ \bibnamefont
  {Garrison}}, \bibinfo {author} {\bibfnamefont {S.}~\bibnamefont {Bieri}}, \
  and\ \bibinfo {author} {\bibfnamefont {C.}~\bibnamefont {Xu}},\ }\href
  {\doibase 10.1103/PhysRevLett.111.157203} {\bibfield  {journal} {\bibinfo
  {journal} {Phys. Rev. Lett.}\ }\textbf {\bibinfo {volume} {111}},\ \bibinfo
  {pages} {157203} (\bibinfo {year} {2013})}\BibitemShut {NoStop}%
\bibitem [{\citenamefont {Grover}\ \emph {et~al.}(2010)\citenamefont {Grover},
  \citenamefont {Trivedi}, \citenamefont {Senthil},\ and\ \citenamefont
  {Lee}}]{grover10}%
  \BibitemOpen
  \bibfield  {author} {\bibinfo {author} {\bibfnamefont {T.}~\bibnamefont
  {Grover}}, \bibinfo {author} {\bibfnamefont {N.}~\bibnamefont {Trivedi}},
  \bibinfo {author} {\bibfnamefont {T.}~\bibnamefont {Senthil}}, \ and\
  \bibinfo {author} {\bibfnamefont {P.~A.}\ \bibnamefont {Lee}},\ }\href
  {\doibase 10.1103/PhysRevB.81.245121} {\bibfield  {journal} {\bibinfo
  {journal} {Phys. Rev. B}\ }\textbf {\bibinfo {volume} {81}},\ \bibinfo
  {pages} {245121} (\bibinfo {year} {2010})}\BibitemShut {NoStop}%
\end{thebibliography}%

\end{document}